\pdfoutput=1
\documentclass[11pt,twoside,a4paper,cmspaper,final,collab]{cms-tdr}
\def\svnVersion{f90d49a}\def\svnDate{2021/10/27}\def\cmsCernNoTag{CERN-EP-2021-085}\def\cmsCernDate{\today}\def\cmsMessage{Published in the Journal of High Energy Physics as \href{http://dx.doi.org/10.1007/JHEP11(2021)225}{\doi{10.1007/JHEP11(2021)225}.}}
\begin{document}\cmsNoteHeader{BPH-18-003}

\pdfoutput=1
\newlength\cmsTabSkip\setlength{\cmsTabSkip}{1ex}
\newcommand{\pis}{\ensuremath{\pi^+_{\mathrm{s}}}\xspace}
\providecommand{\cmsTable}[1]{\resizebox{\textwidth}{!}{#1}}

\title{Measurement of prompt open-charm production cross sections in proton-proton collisions at \texorpdfstring{$\sqrt{s} = 13\TeV$}{sqrt(s) = 13 TeV}}

\abstract{
The production cross sections for prompt open-charm mesons in proton-proton collisions at a center-of-mass energy of 13\TeV are reported. 
The measurement is performed using a data sample collected by the CMS experiment corresponding to an integrated luminosity of 29\nbinv.
The differential production cross sections of the \PDstpm, \PDpm, and \PDz (\PaDz) mesons are presented in ranges of transverse momentum and pseudorapidity $4<\pt<100\GeV$ and $\abs{\eta}<2.1$, respectively.
The results are compared to several theoretical calculations and to previous measurements.}

\hypersetup{
pdfauthor={CMS Collaboration},
pdftitle={Measurement of prompt open-charm production cross sections in proton-proton collisions at sqrt(s) = 13 TeV},
pdfsubject={CMS},
pdfkeywords={CMS, charm, B physics}}
\maketitle

\maketitle
\flushbottom

\section{Introduction}
\label{intro}
Measurements of the production cross sections for open-charm mesons, \ie, mesons containing a single charm quark, in hadronic collisions at LHC center-of-mass energies provide an important test of the theory of strong interactions known as quantum chromodynamics (QCD).
Due in part to the presence of several competing scales (charm mass, transverse momentum) that are close to the threshold for the validity of the perturbative expansion, the theoretical uncertainties are rather large. 
Therefore, experimental constraints on heavy-quark production cross sections are relevant for physics phenomena directly involving heavy quarks or for which they constitute a background. 
An improved understanding of charm production is furthermore relevant for the exploration of physics processes arising from charm hadron decays, such as neutrinos and other feebly interacting particles \cite{faser, ship, snd}, and as a reference for the study of the properties of QCD extreme media \cite{cmsHN_hep}.
Several studies have been carried out at the LHC. These include measurements at $\sqrt{s} = 7\TeV$ by the ATLAS Collaboration \cite{atlas}, at $\sqrt{s} = 5$ \cite{alice_5, alice_5_charm_beauty}, 7 \cite{alice, alice2}, and 13\TeV \cite{alice_13} by the ALICE experiment, and at $\sqrt{s} = 5$ \cite{lhcb_5}, 7 \cite{lhcb_7}, and 13\TeV \cite{lhcb_13} by the LHCb Collaboration.
The CMS experiment has produced results on open-charm production, analyzing heavy-ion and proton-proton (\Pp{}\Pp) collisions at $\sqrtsNN = 5.02\TeV$ \cite{cmsHN_hep}.

In this paper, a study of charm meson production by CMS in \Pp{}\Pp collisions at $\sqrt{s} = 13\TeV$ is presented.
The analysis is focused on the measurement of cross sections for the prompt production of \PDstp, \PDz, and \PDp mesons.
The charm mesons are identified via their exclusive decays:
\begin{itemize}
\item $\Pp{}\Pp \to \PDstp X \to \PDz \pis X \to \PKm\Pgpp\pis X$,
\item $\Pp{}\Pp \to \PDz X \to \PKm\Pgpp X$,
\item $\Pp{}\Pp \to \PDp X \to \PKm\Pgpp\Pgpp X$,
\end{itemize}
where $X$ corresponds to any set of possible particles. 
Charge conjugation is implied throughout the paper, unless specified otherwise.
The designation \pis indicates that, because of the specific kinematics of the \PDstp decay, this "slow" pion has significantly lower momentum than the kaon and pion decay products of the \PDz.
The \PDstp, \PDz, and \PDp mesons are reconstructed in the range of transverse momentum $4<\pt<100\GeV$ and pseudorapidity $\abs{\eta}<2.1$.

Charm mesons arising from the \Pp{}\Pp collision point, either directly or as decay products of excited charm resonances (\eg, \PDz coming from \PDstp decay), are referred to as promptly produced.
Contrarily, charm mesons coming from the decays of \Pb hadrons, \eg, $\PB\to \PD X$, where $X$ denotes any additional particles, are referred to as nonprompt and are considered as a background process. 

Since the kinematic regions covered by the existing measurements differ from the one presented in this paper and are fully complementary in the case of LHCb, the new measurements described in this paper and their comparison to theoretical predictions provide an important contribution to a deeper understanding of the charm production mechanism.

The single-differential cross sections for prompt charm meson production are measured as a function of the transverse momentum \pt and the absolute value of pseudorapidity $\abs{\eta}$. 
In principle, rapidity ($y$) is the proper kinematic variable to study cross sections of massive particles. 
Here, pseudorapidity is used instead of rapidity to facilitate the comparison with the ATLAS measurement \cite{atlas}, which is the closest in kinematic range to the results presented in this paper.
In the kinematic phase space of this measurement, the maximum difference between $\eta$ and $y$ is less than 5\%.
Tabulated results are provided in HEPData~\cite{hep_data}.
The measured cross sections are compared to the predictions from the Monte Carlo (MC) event generators \PYTHIA 6.4 \cite{pythia6} and 8.1 \cite{pythia8}, to the theoretical calculations from FONLL \cite{fonll, fonll2}, and to the previous LHC results \cite{atlas, alice_5, alice, alice2, lhcb_5, lhcb_7, lhcb_13, cmsHN_hep}. 

\section{The CMS detector}
\label{cms}

The central feature of the CMS apparatus is a superconducting solenoid of 6\unit{m} internal diameter, providing a magnetic field of 3.8\unit{T}. 
Within the solenoid volume are a silicon pixel and strip tracker, a lead tungstate crystal electromagnetic calorimeter, and a brass and scintillator hadron calorimeter, each composed of a barrel and two endcap sections. 
Forward calorimeters extend the pseudorapidity coverage provided by the barrel and endcap detectors. 
Muons are detected in gas-ionization chambers embedded in the steel flux-return yoke outside the solenoid.
The most relevant subdetector for this analysis is the silicon tracker.
The silicon tracker measures charged particles within the pseudorapidity range $\abs{\eta} < 2.5$. It consists of 1440 silicon pixel and 15\,148 silicon strip detector modules. For nonisolated particles of $1 < \pt < 10\GeV$ and $\abs{\eta} < 1.4$, the track resolutions are typically 1.5\% in \pt and 25--90 (45--150)\mum in the transverse (longitudinal) impact parameters \cite{TRK-11-001}.

Events of interest are selected using a two-tiered trigger system~\cite{Khachatryan:2016bia}. 
The first level, composed of custom hardware processors, uses information from the calorimeters and muon detectors to select events at a rate of around 100\unit{kHz} within a time interval of less than 4\mus. 
The second level, known as the high-level trigger, consists of a farm of processors running a version of the full event reconstruction software optimized for fast processing, and reduces the event rate to around 1\unit{kHz} before data storage.
A more detailed description of the CMS detector, together with a definition of the coordinate system used and the relevant kinematic variables, can be found in Ref. \cite{cms}.

\section{Data samples, simulation, and event selection}
\label{data_simulation_event}

The data sample was acquired in 2016 and corresponds to an integrated luminosity of 29\nbinv, out of the 36.8\fbinv collected in that year \cite{lumi_last}. 
The average number of simultaneous \Pp{}\Pp collisions in the same or nearby bunch crossings, referred to as pileup (PU), for the subset of data used for this analysis is 14.
To avoid any bias from a trigger requirement that would aim to select events and require an efficiency correction, the data used in this analysis were collected with an unbiased trigger that only required the presence of crossing beams.
This trigger was  heavily prescaled, which explains the low effective integrated luminosity for  this analysis with respect to the total one collected in 2016.

The candidate vertex with the largest value of summed physics object $\pt^2$ is taken to be the primary \Pp{}\Pp interaction vertex (PV). 
The physics objects are the jets, clustered using the jet-finding algorithm~\cite{Cacciari:2008gp,Cacciari:2011ma} with the tracks assigned to candidate vertices as inputs, and the associated missing transverse momentum, taken as the negative vector sum of the $\vec{\pt}$ of those jets. 

The effects arising from the detector acceptance, reconstruction efficiency, and selection efficiency, whose combination is referred to as the total reconstruction efficiency, are determined from simulated events. 
These events are generated with \PYTHIA6.4 \cite{pythia6}, the heavy-flavor hadrons are decayed with \EVTGEN~1.3.0 \cite{evtgen}, and the final-state particles are propagated through a simulation of the CMS detector based on \GEANTfour v10.00.p02 \cite{geant}. 
The simulated events used to determine the efficiency demand a \PDstp meson with $\pt>3.9\GeV$, which is required to decay via $\PDstp \to \PDz\pis \to \PKm \Pgpp \pis$. 
The \pt threshold does not bias the \pt spectrum for the \PDstp and \PDz events as this measurement is only for \PD mesons with $\pt>4\GeV$. 
This sample is also used to determine the \PDp efficiency, where the \PDp originates from the hadronization of the other charm quark in each event. 
In this case, the \pt spectrum is biased by the \pt threshold on the \PDstp, and therefore, the simulated \PDp \pt distribution is reweighted to match the \PDstp spectrum. 
The effects of PU have been included by overlaying generated minimum-bias events with the main simulated collision. 
The distribution of the number of PU events is reweighted in the simulation to match the observed data distribution, separately for the 7 main data-taking periods. 
Following these corrections, the distributions of the kinematic and selection criteria variables obtained from simulation are found to agree with the data for all three mesons.
 
\section{Analysis Strategy}
   
\subsection{Charm meson reconstruction}
\label{reco}

The first step in the reconstruction of the charm mesons is the selection of tracks corresponding to the final-state objects.
The criteria used to select the tracks include a minimum \pt, a maximum $\chi^2$ of the track fit divided by the number of degrees of freedom (dof), a minimum number of hits in the pixel detector and in the full tracker (pixel and strip detectors), and maximum impact parameters with respect to the PV in the transverse plane ($\mathrm{IP}_{xy}$)  and longitudinal direction ($\mathrm{IP}_z$).  
The track requirements are summarized in Table \ref{cuts}.

\begin{table}[ht!]
\topcaption{The selection requirements for each charm meson.} 
\centering
\begin{tabular}{llll}

 Variables & \PDstp & \PDz & \PDp \\
 \hline
 Tracks: \pt (\GeVns)  & ${>}\,0.5$ (${>}\,0.3$ for \pis) & ${>}\,0.8$ & ${>}\,0.7$ \\
 Tracks: $\chi^2/\mathrm{dof}$ & ${<}\,2.5$ (${<}\,3$ for \pis) & ${<}\,2.5$ & ${<}\,2.5$ \\
 Tracks: tracker hits & ${\ge}\,5$ (${\ge}\,3$ for \pis) & ${\ge}\,5$ & ${\ge}\,5$ \\
 Tracks: pixel hits & ${\ge}\,2$ (none for \pis) & ${\ge}\,2$ & ${\ge}\,2$  \\
 Tracks: IP$_{xy}$ (cm) & ${<}\,0.1 $ (${<}\,3\sigma$ for \pis) & ${<}\,0.1 $ & ${<}\,0.1 $ \\
 Tracks: IP$_z$ (cm) & ${<}\,1$ (${<}\,3\sigma$ for \pis) & ${<}\,1 $ & ${<}\,1 $ \\
 $\abs{M_{\text{cand}} - M^{\text{PDG}}}$ (\GeVns) & ${<}\,0.023$ (for \PDz) & ${<}\,0.1$ & ${<}\,0.1$ \\
 SV fit probability & \multicolumn{3}{c}{${>}\,1\%$} \\
 Pointing, $\cos\alpha$ & \multicolumn{3}{c}{${>}\,0.99$} \\
 SV significance & ${>}\,3$  & ${>}\,5$  & ${>}\,10$  \\
 
\end{tabular}
\label{cuts}
\end{table}

The \PDz (\PDp) mesons are reconstructed by combining two (three) tracks with total charge 0 (1) and having an invariant mass $M_{\text{cand}}$ within 100\MeV of the nominal meson mass $M^\text{PDG}$~\cite{pdg}.
When the \PDz candidates are reconstructed in the \PDstp decay chain, a mass window of 23\MeV on the \PDz mass is used, instead.
In the \pt range relevant for this analysis, charged pions and kaons cannot be identified efficiently in the CMS detector.
A kaon or pion mass hypothesis is thus assumed for the tracks, according to the charge and the specific decay channel. 
Three topological requirements, whose values are given in Table \ref{cuts}, are also used to reduce the background, which is primarily from random combinations of tracks.  
First, the secondary vertex (SV) fit $\chi^2$ probability from fitting the two (\PDz) or three (\PDp) tracks to a common vertex is used to ensure the tracks originate from a common point. 
Second, the cosine of the angle ($\cos\alpha$) between the charm candidate momentum and the vector pointing from the PV to the SV is used to ensure the \PD meson is consistent with originating from the PV, which reduces background from \Pb hadron decays as well as from random track combinations. 
Third, the SV significance is the distance between the PV and SV divided by its uncertainty. 
This is a crucial requirement in the analysis, which provides a considerable reduction in the combinatorial background that is dominated by tracks originating from the PV.

To complete the \PDstp meson reconstruction, a third track, corresponding to the slow pion, has a pion mass assigned and is kinematically combined with the \PDz candidate. 
Looser requirements on the \pt, $\chi^2$/dof, and total number of hits are used for this track, as detailed in Table \ref{cuts}. 
In addition, the impact parameter requirements are changed to require that $\mathrm{IP}_{xy}$ and $\mathrm{IP}_{z}$ be less than three times their respective uncertainties. 
To improve the mass resolution, the mass difference $\Delta M = m(\PK\pi\pis) - m(\PK\pi)$ is used in the analysis instead of the invariant mass of the three-track combination.

For each event, we require there be no more than one candidate for each of the six decay modes (three mesons and two charge-conjugate states). 
For events in which there is more than one candidate in a particular decay mode, the candidate whose invariant mass is closest to $M^{PDG}(\PD)$~\cite{pdg} is chosen for \PDz and \PDp candidates and the smallest $\Delta M$ for the \PDstp candidates. 
This arbitration is required for 2, 3, and 11$\%$ of the events with a \PDstp, \PDz, and \PDp candidate, respectively. 
By comparing with a random arbitration scheme, it was verified that this method does not introduce a statistically  significant bias in terms of the signal yield or signal invariant mass distribution. 
However, a bias in the background shape has been identified for the \PDp mesons, which is described in Section \ref{syst}.

\subsection{Signal yield determination}
\label{yield}

The prompt charm meson differential cross section $\rd\sigma/\rd\pt$ is measured in 9 bins of \pt between 4 and 100\GeV in the range $\abs{\eta} < 2.1$; 
the differential cross section $\rd\sigma/\rd\abs{\eta}$ is measured in 10 bins of $\abs{\eta}$, for $\abs{\eta}< 2.1$ and $4<\pt<100\GeV$.

The signal yields, including both prompt and nonprompt decays, are determined using unbinned maximum-likelihood fits to the invariant mass distributions for the various decay modes (the $\Delta M$ distribution is used for the \PDstp) in each \pt and $\abs{\eta}$ bin.
The signal components are modeled by the sum of two Gaussian functions to account for the nonuniform resolution over the detector acceptance.
The two Gaussian function means are constrained to be the same.
The mean, widths, and normalizations are treated as free parameters.
An additional Gaussian function is used to describe the invariant mass shape of \PDz candidates with incorrect pion and kaon mass assignments.
The width of this wide Gaussian is taken from simulation bin by bin.
The normalization of the wide Gaussian function contribution is fixed to be the same as that of the sum of the two narrow signal Gaussian functions in each bin, reflecting the fact that the number of correct and swapped \PK/\Pgp \PDz signal candidates is the same by construction.

The combinatorial background is described with different functions, according to the decay mode.
For the \PDstp meson, the background is described by a phenomenological threshold function \cite{trk_2010} given by
\begin{equation}
f = \left( 1-\re^{ - { \frac{\Delta M - M_0}{p_0} } } \right) \left( { \frac{\Delta M }{M_0} } \right) ^{p_1} + p_2 \left( { \frac{\Delta M}{M_0} } -1 \right),
\label{comb_funct}
\end{equation}
where $M_0$ is the endpoint, taken to be the pion mass, and $p_{0,1,2}$ are free parameters. 
For the \PDz and \PDp mesons, the combinatorial background component is modeled by a third-degree polynomial function, where the four parameters, including the normalization, are unconstrained in the fits.

\begin{figure}[ht!]
\centering
\includegraphics[width=0.4\textwidth]{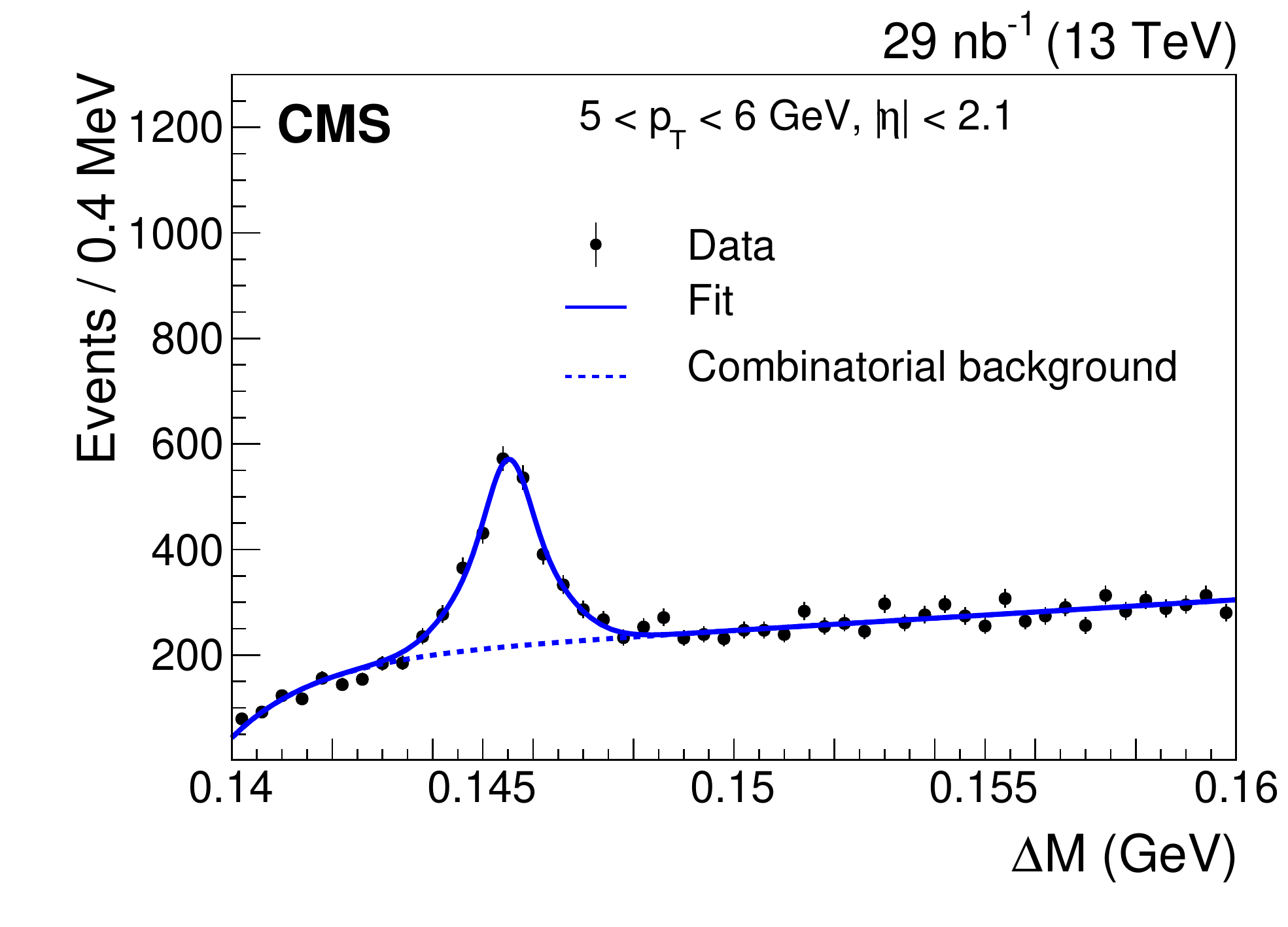}
\includegraphics[width=0.4\textwidth]{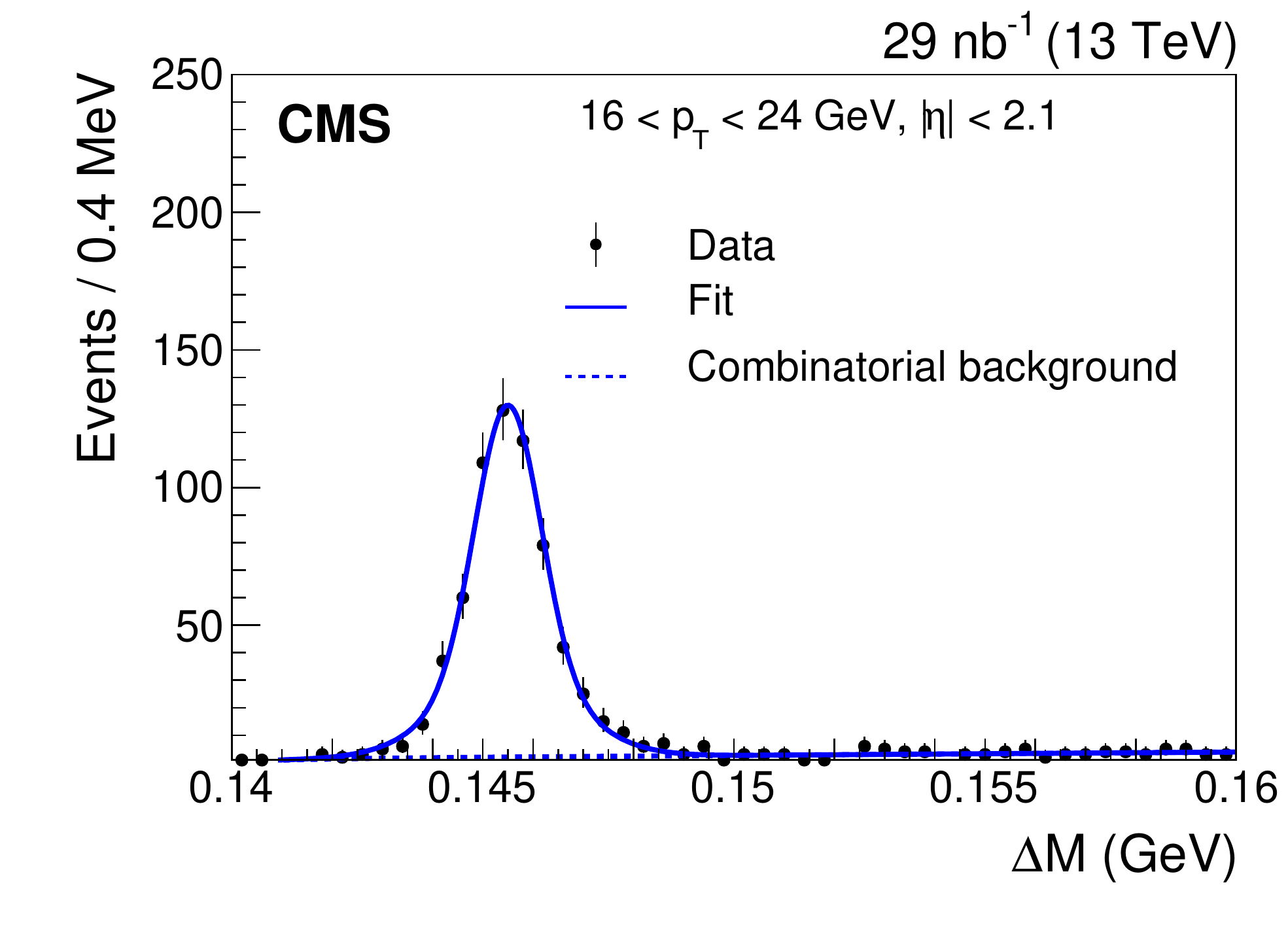} \\
\includegraphics[width=0.4\textwidth]{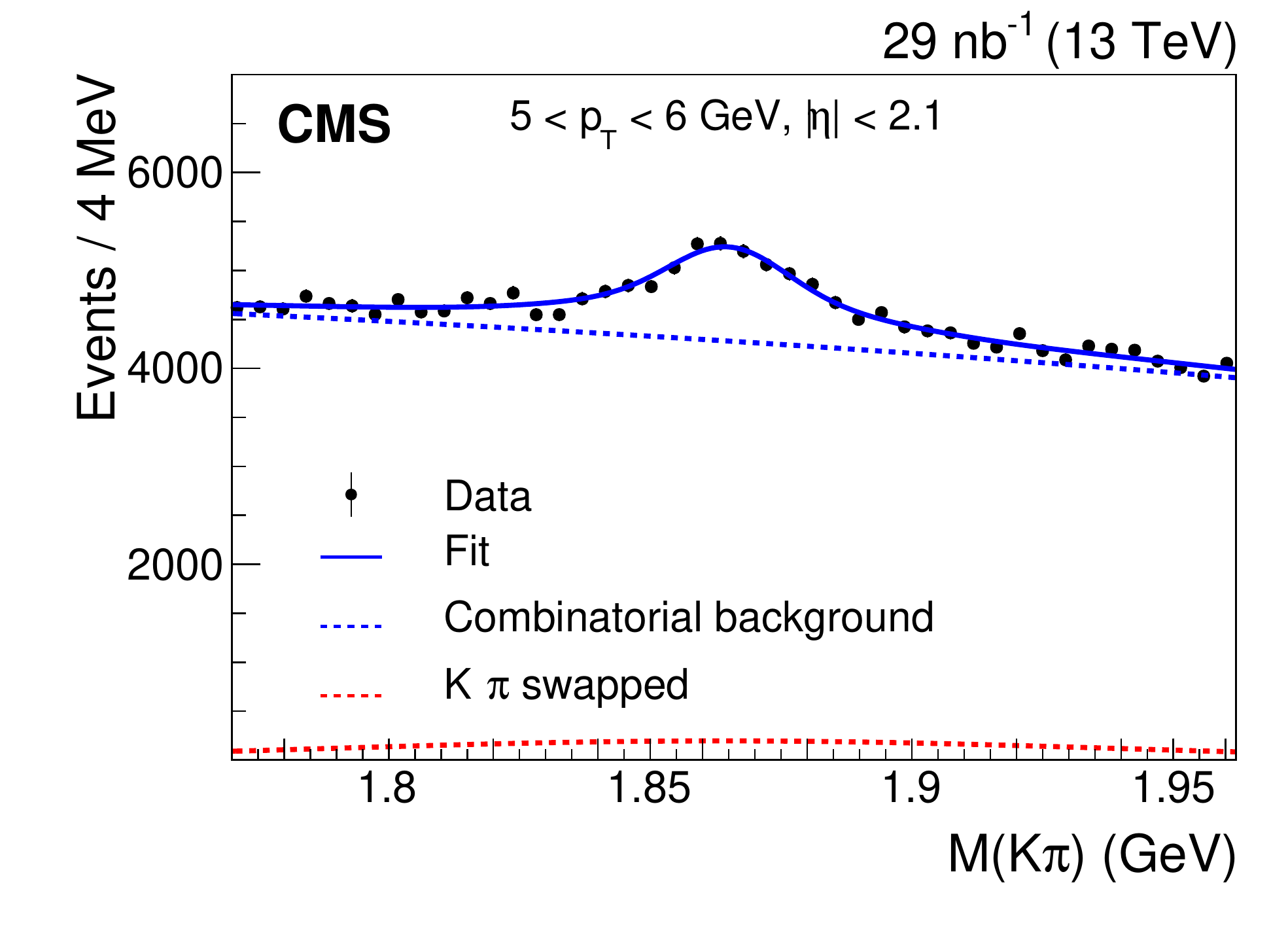}
\includegraphics[width=0.4\textwidth]{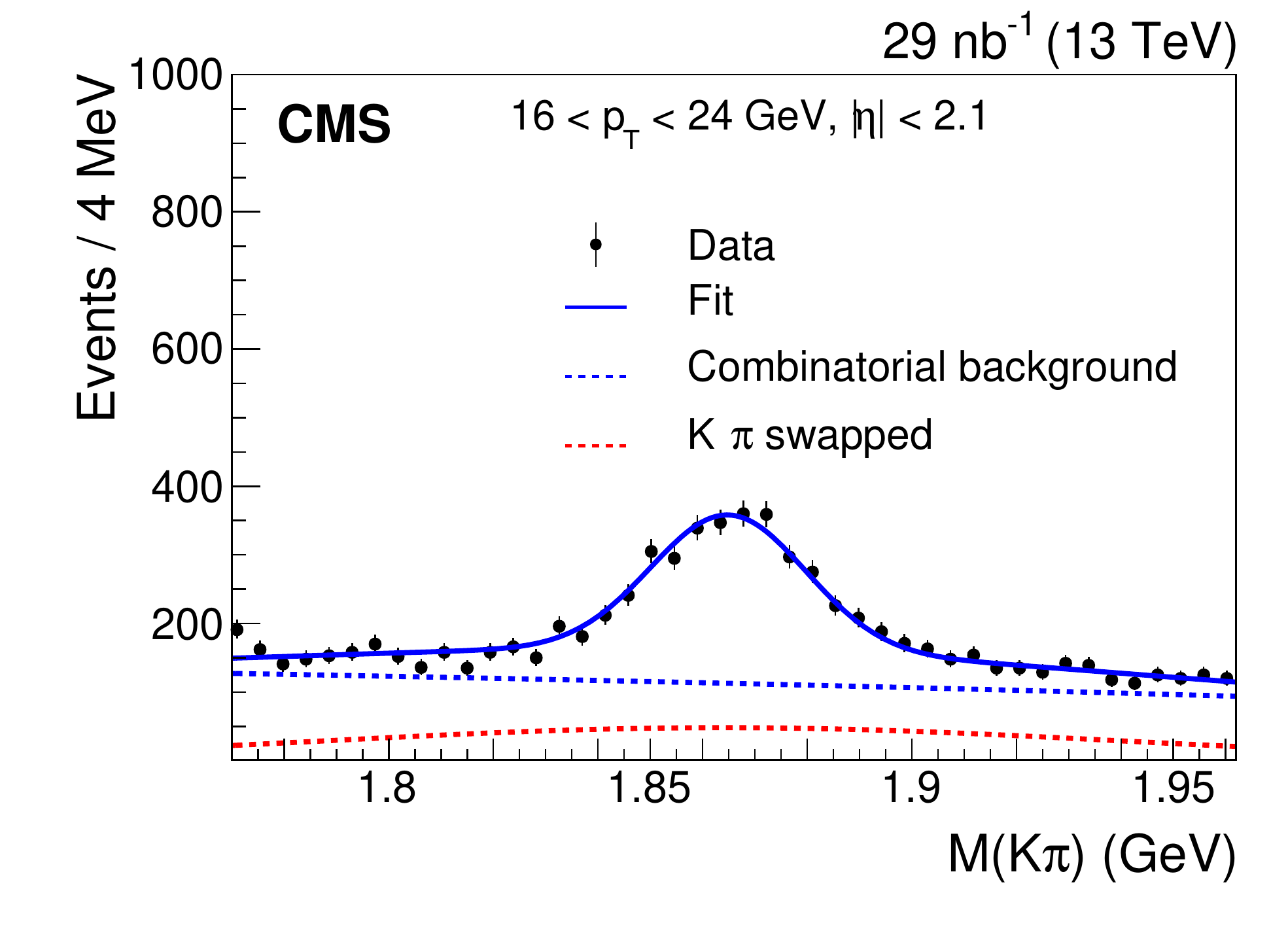}\\
\includegraphics[width=0.4\textwidth]{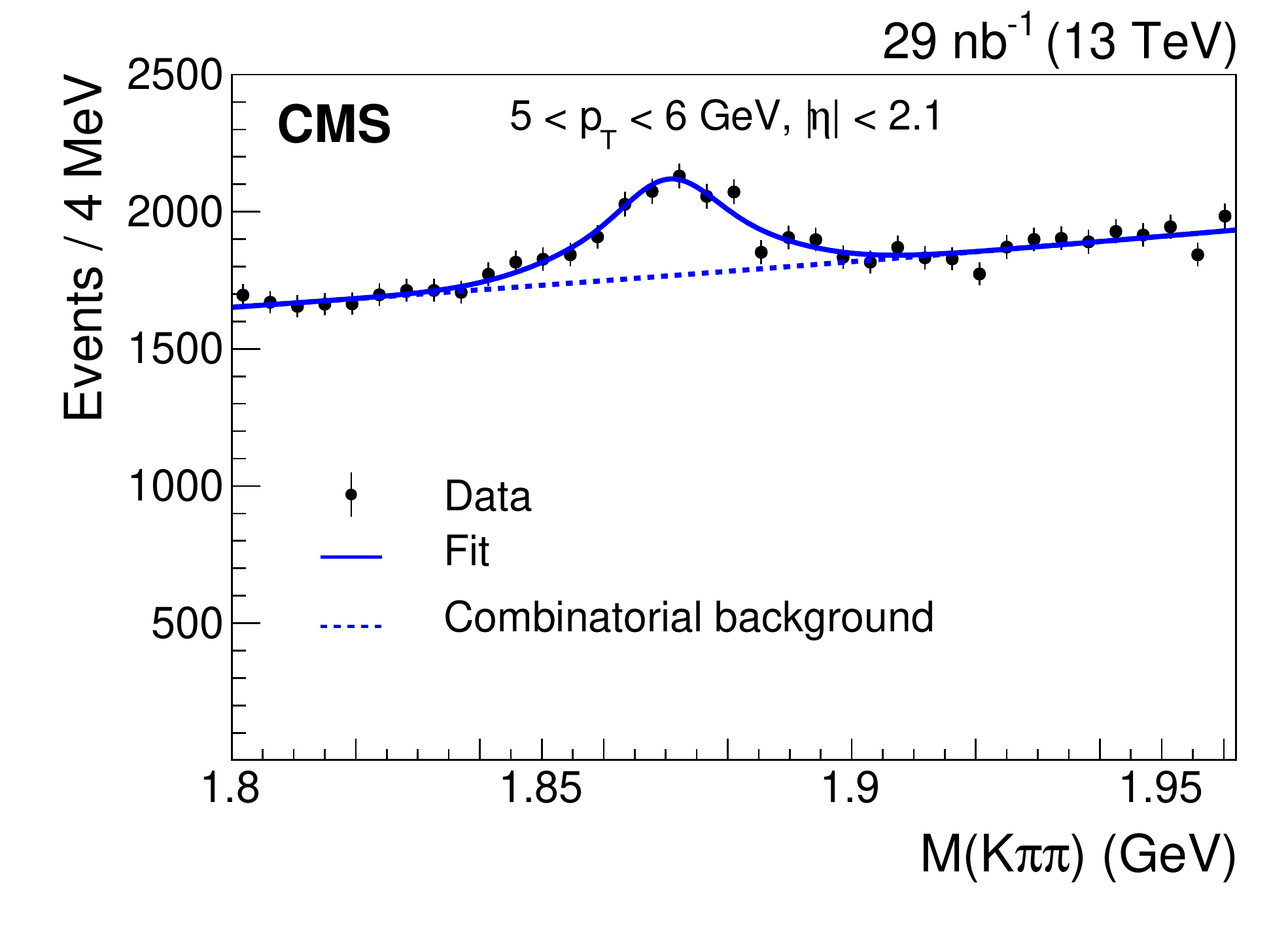}
\includegraphics[width=0.4\textwidth]{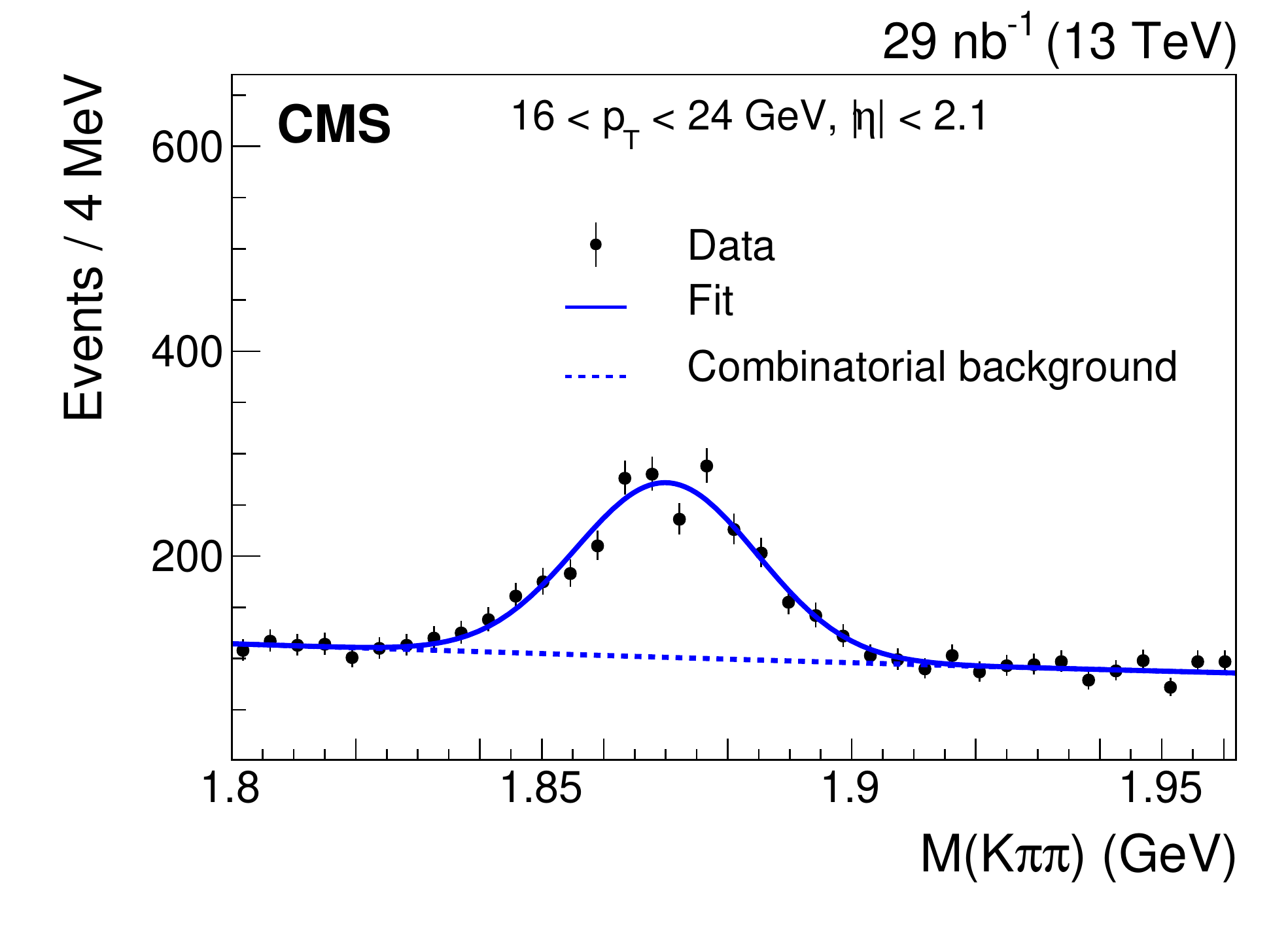}\\
\caption{The invariant mass distributions: $\Delta M = M(\PKm\Pgpp\pis) - M(\PKm\Pgpp)$ (upper), $M(\PKm\Pgpp)$ (middle), and $M(\PKm\Pgpp\Pgpp)$ (lower); charge conjugation is implied. 
Plots in the left column show the $5 < \pt < 6\GeV$ bin, while the $16 < \pt < 24\GeV$ bin is shown in the right column. 
The vertical bars on the points represent the statistical uncertainties in the data. 
The overall result from the fit is shown by the solid line, the fit to the combinatorial background by the dotted line, and, in the middle plots, the fit to the \PK/\Pgp swapped candidates by the red dot-dashed line.
} 
\label{Dmesons_fit_pt}
\end{figure}

\begin{figure}[ht!]
\centering
\includegraphics[width=0.4\textwidth]{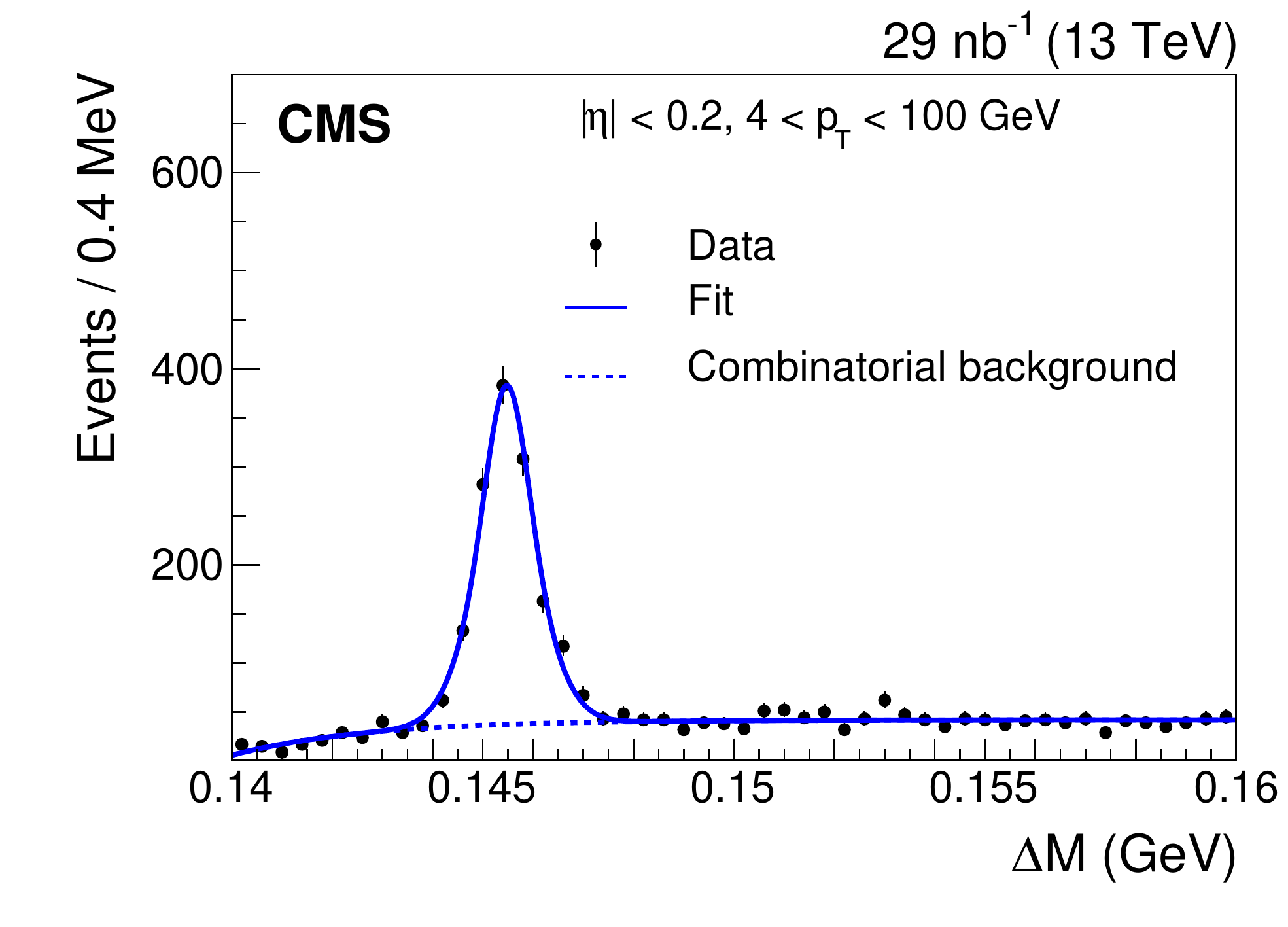}
\includegraphics[width=0.4\textwidth]{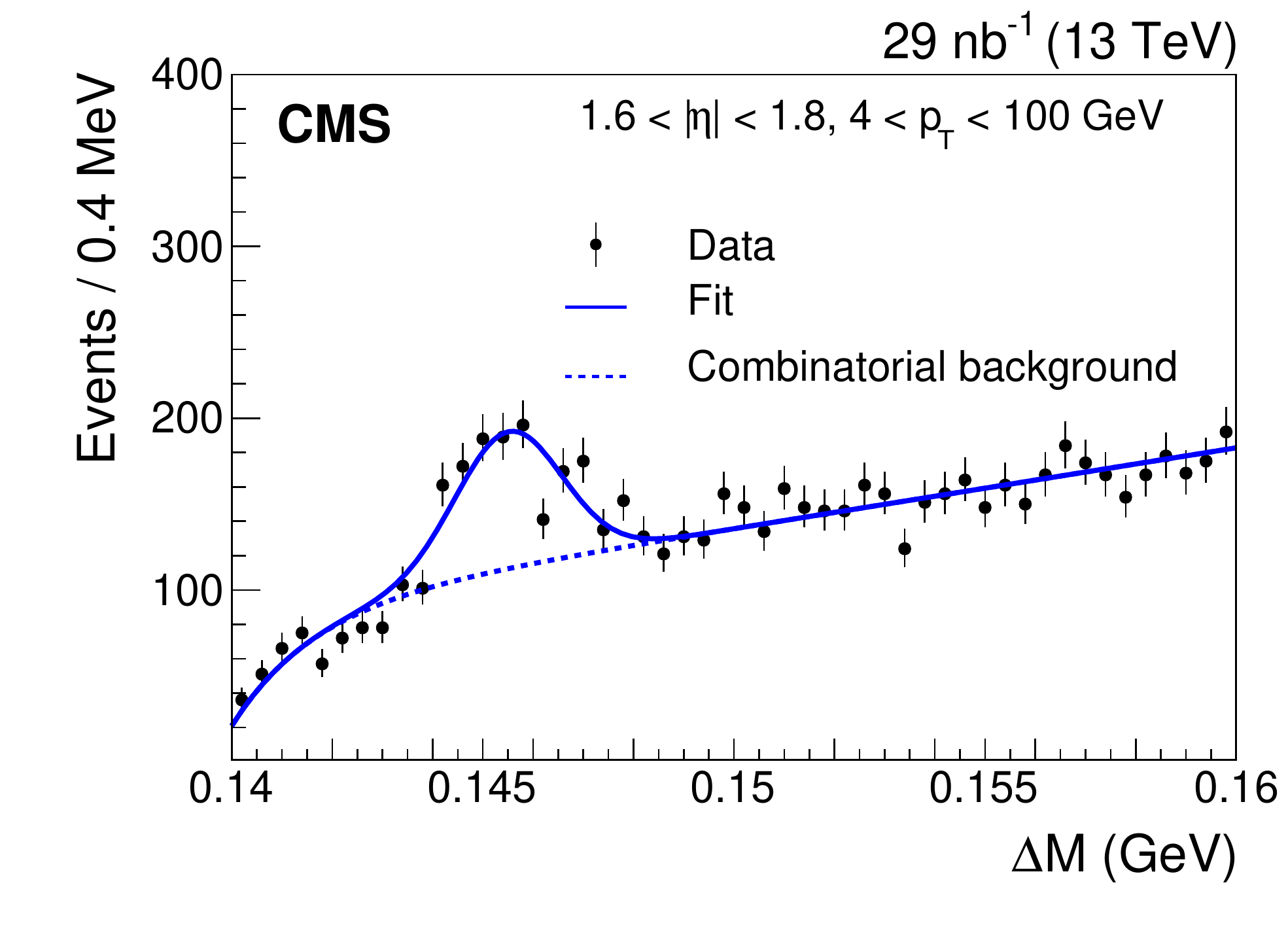} \\
\includegraphics[width=0.4\textwidth]{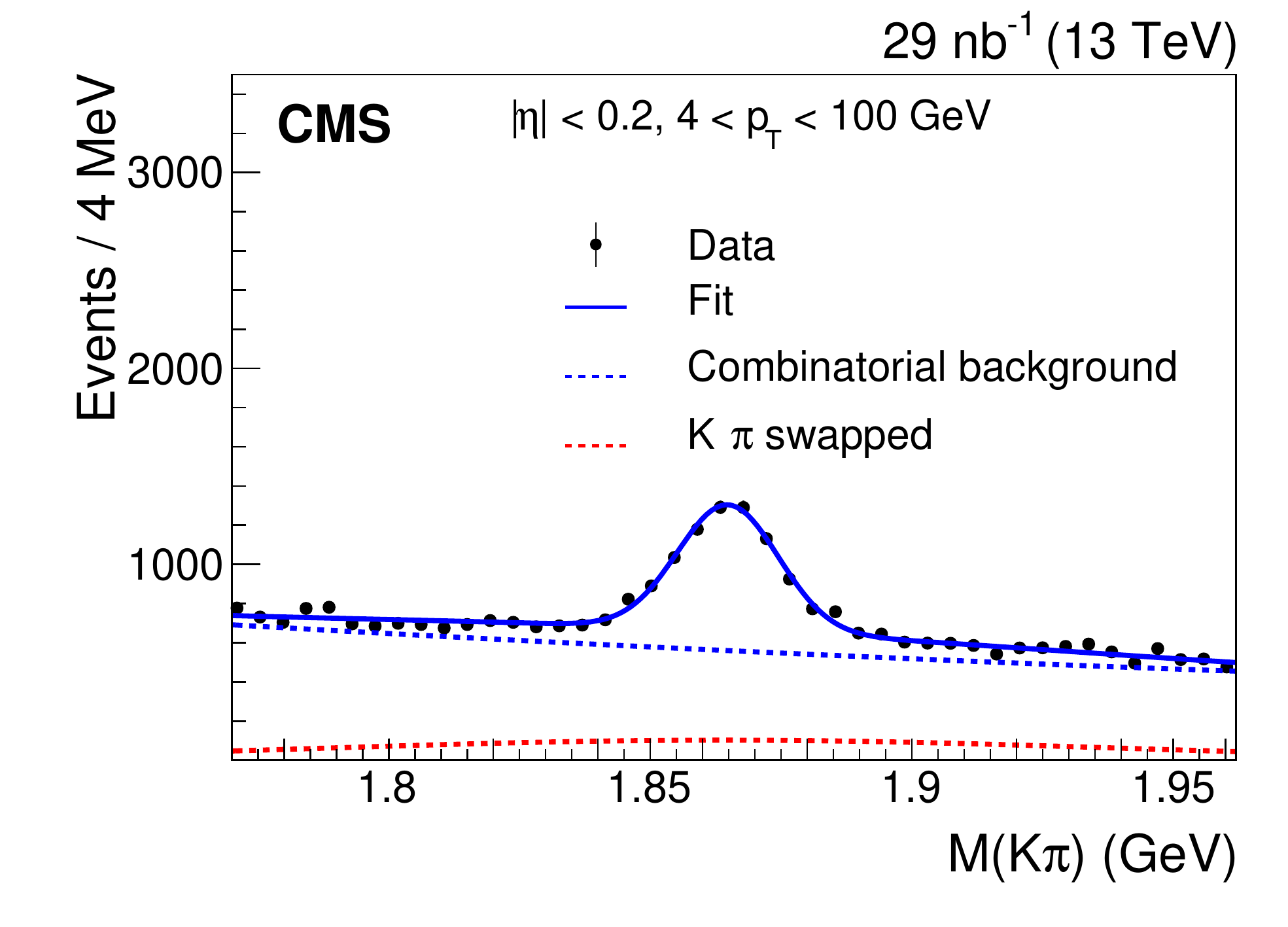}
\includegraphics[width=0.4\textwidth]{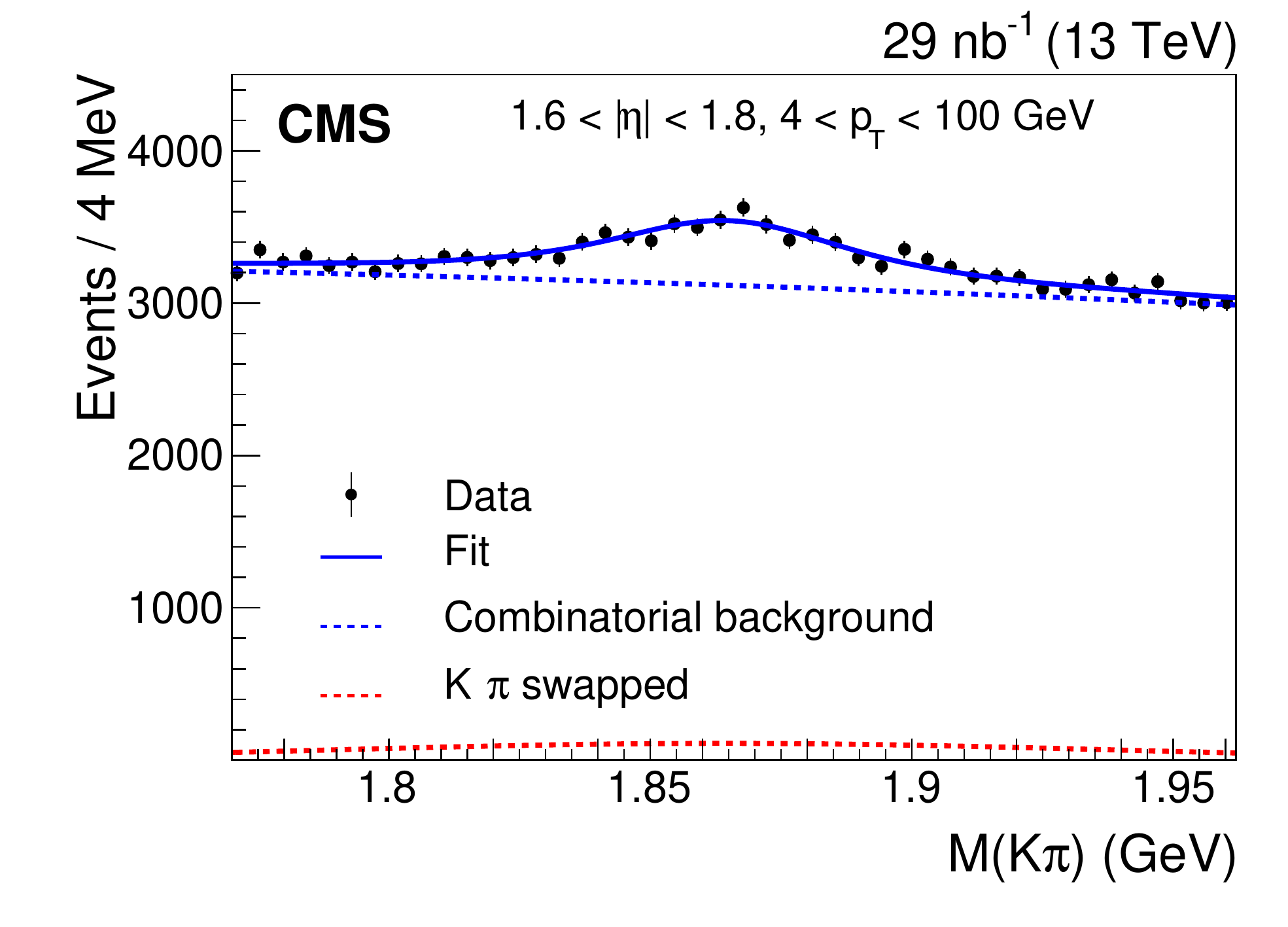}\\
\includegraphics[width=0.4\textwidth]{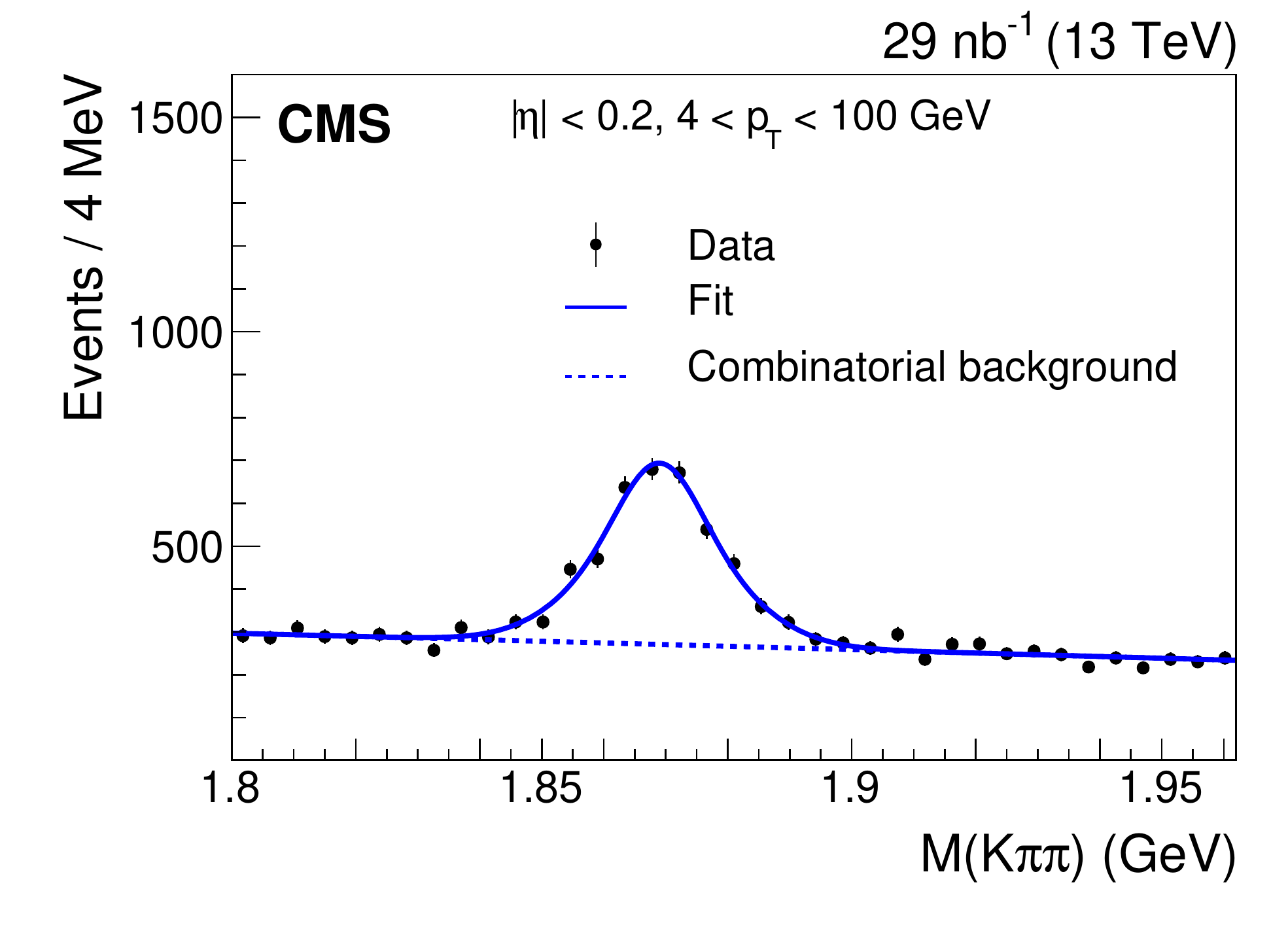}
\includegraphics[width=0.4\textwidth]{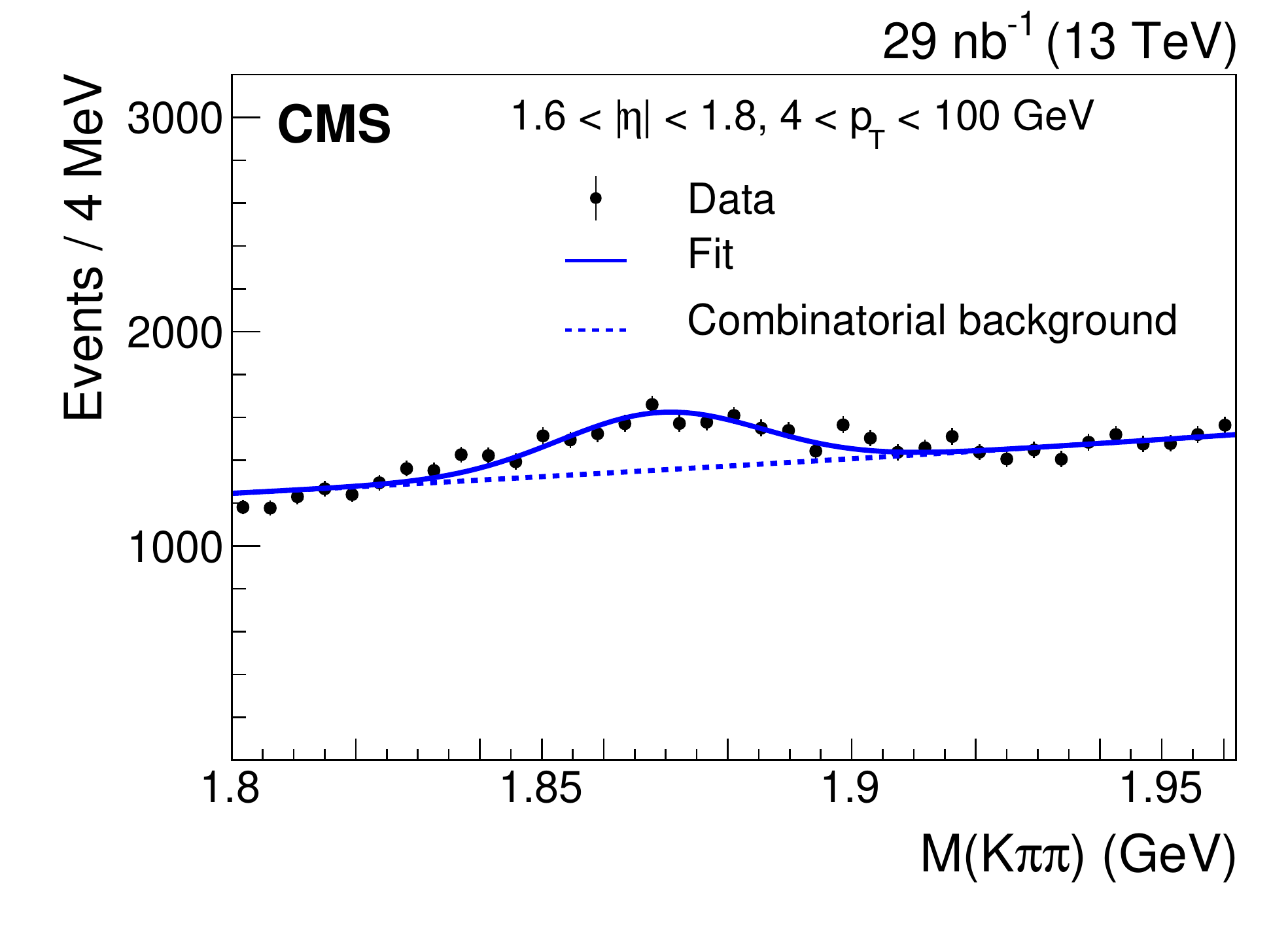}\\
\caption{
The invariant mass distributions: $\Delta M = M(\PKm\Pgpp\pis) - M(\PKm\Pgpp)$ (upper), $M(\PKm\Pgpp)$ (middle), and $M(\PKm\Pgpp\Pgpp)$ (lower); charge conjugation is implied.
Plots in the left column show the $\abs{\eta} < 0.2$ bin, while the $1.6 < \abs{\eta} < 1.8$ bin is shown in the right column. 
The vertical bars on the points represent the statistical uncertainties in the data. 
The overall result from the fit is shown by the solid line, and the fit to the combinatorial background by the dotted line, and, in the middle plots, the fit to the \PK/\Pgp swapped candidates by the red dot-dashed line.}
\label{Dmesons_fit_eta}
\end{figure}

In Fig. \ref{Dmesons_fit_pt}, the fitted invariant mass distributions are reported for two example \pt bins, low (5--6\GeVns) and high (16--24\GeVns); while Fig. \ref{Dmesons_fit_eta} shows the fitted invariant mass distributions for the bins $\abs{\eta}< 0.2$ and $1.6 <\abs{\eta}< 1.8$. 
As expected, at low \pt and high $\abs{\eta}$, the track momentum and position resolutions are worse, which affect the reconstructed mass width and the distribution shapes, resulting in an increase in the combinatorial background under the peak. 
Despite the different kinematic regions, it was found that the same functions with different parameter values reproduce all the distributions well.

The signal yields and the statistical uncertainties returned by the fits are reported in Tables \ref{yield_pt} and \ref{yield_eta} for the \pt and $\abs{\eta}$ bins, respectively.

\begin{table}[ht!]
 \topcaption{The signal yields in data for \PDstp, \PDz, and \PDp mesons in \pt bins for $\abs{\eta} < 2.1$. Only the statistical uncertainties are reported.}
  \centering
  \begin{tabular}{cccc}
   \pt range (\GeVns) & \PDstp & \PDz & \PDp \\ 
      \hline
      \begin{tabular}{r@{--}l}
      4 & 5 \\
      5 & 6 \\ 
      6 & 7 \\ 
      7 & 8 \\
      8 & 12 \\
     12 & 16 \\ 
     16 & 24 \\ 
     24 & 40 \\ 
     40 & 100 \\
      \end{tabular}
     & 
     \begin{tabular}{r@{ $\pm$ }l}
      901 & 40 \\ 
      1555 & 99 \\ 
      1450 & 98 \\
      1144 & 75 \\
      2962 & 80 \\
      1170 & 63 \\
      647 & 28 \\
      180 & 15 \\
      38 & 6\\ 
      \end{tabular} 
      &
      \begin{tabular}{r@{ $\pm$ }l}
      5200 & 700 \\
      5480 & 400 \\
      5640 & 450 \\ 
      3990 & 500 \\
      9750 & 320 \\ 
      3540 & 170 \\
      2040 & 120 \\ 
      625 & 58 \\
      91 & 20 \\
      \end{tabular}
      & 
      \begin{tabular}{r@{ $\pm$ }l}
      5850 & 1400 \\
      3390 & 460 \\ 
      3620 & 140 \\
      2790 & 130 \\
      6860 & 180 \\
      3047 & 96 \\ 
      1821 & 77 \\
      628 & 61 \\
      83 & 11 \\

       \end{tabular}
  \end{tabular}
  \label{yield_pt}
\end{table}

\begin{table}
 \topcaption{The signal yields in data for \PDstp, \PDz, and \PDp mesons with $ 4<\pt<100\GeV$ in $\abs{\eta}$ bins. Only the statistical uncertainties are reported.}
  \centering
  \begin{tabular}{cccccc}
   $\abs{\eta}$ range & \PDstp & \PDz &  \PDp \\ 
   \hline
     \begin{tabular}{r@{--}l}
      0 & 0.2 \\
      0.2 & 0.4 \\
      0.4 & 0.6 \\ 
      0.6 & 0.8 \\ 
      0.8 & 1.0 \\  
      1.0 & 1.2 \\
      1.2 & 1.4 \\
      1.4 & 1.6 \\
      1.6 & 1.8 \\
      1.8 & 2.1 \\ 
     \end{tabular}
     &
     \begin{tabular}{r@{ $\pm$ }l}
     1227 & 42\\
     1340 & 61 \\
     1276 & 45 \\
     1288 & 48 \\ 
     1283 & 68 \\
     1026 & 88 \\ 
     857 & 61 \\
     576 & 59 \\ 
     530 & 84 \\ 
     539 & 69 \\  
     \end{tabular}
     & 
     \begin{tabular}{r@{ $\pm$ }l}
     3640 & 120 \\ 
     4150 & 140 \\ 
     4280 & 150 \\ 
     4590 & 180 \\ 
     3700 & 210 \\ 
     4270 & 320 \\ 
     3180 & 300 \\ 
     3050 & 320 \\ 
     3000 & 460 \\
     3770 & 590 \\ 
     \end{tabular}
     & 
     \begin{tabular}{r@{ $\pm$ }l}
     3200 & 110 \\
     3060 & 96 \\ 
     2910 & 190 \\
     3200 & 230 \\
     3040 & 140 \\ 
     3180 & 160 \\ 
     2850 & 160 \\ 
     2350 & 250 \\
     2130 & 210 \\ 
     2780 & 490 \\
     \end{tabular} 
  \end{tabular}
  \label{yield_eta}
\end{table}  

\subsection{Efficiency estimation}
\label{ch:effi}

The efficiency is estimated using the signal MC sample and is defined as the fraction of charm signal decays, generated in the kinematic region $4 < \pt < 100\GeV$ and $\abs{\eta} < 2.1$, that is reconstructed and survives the selection criteria described in Section \ref{reco}.   
The efficiency is thus determined for each \pt and $\abs{\eta}$ bin and for both the charge-conjugate states.
Taking the \PDstp channel as an example, these values range from 0.6\% for $4 < \pt < 5\GeV$ to 30\% for $40 < \pt < 100\GeV$, and from 3.8\% for $\abs{\eta} < 0.2$ to 1.5\% for $1.8 < \abs{\eta} < 2.1$. 

\subsection{Contamination from nonprompt decays}
\label{contamination}

\begin{figure}[ht!]
\centering
\includegraphics[width=0.47\textwidth]{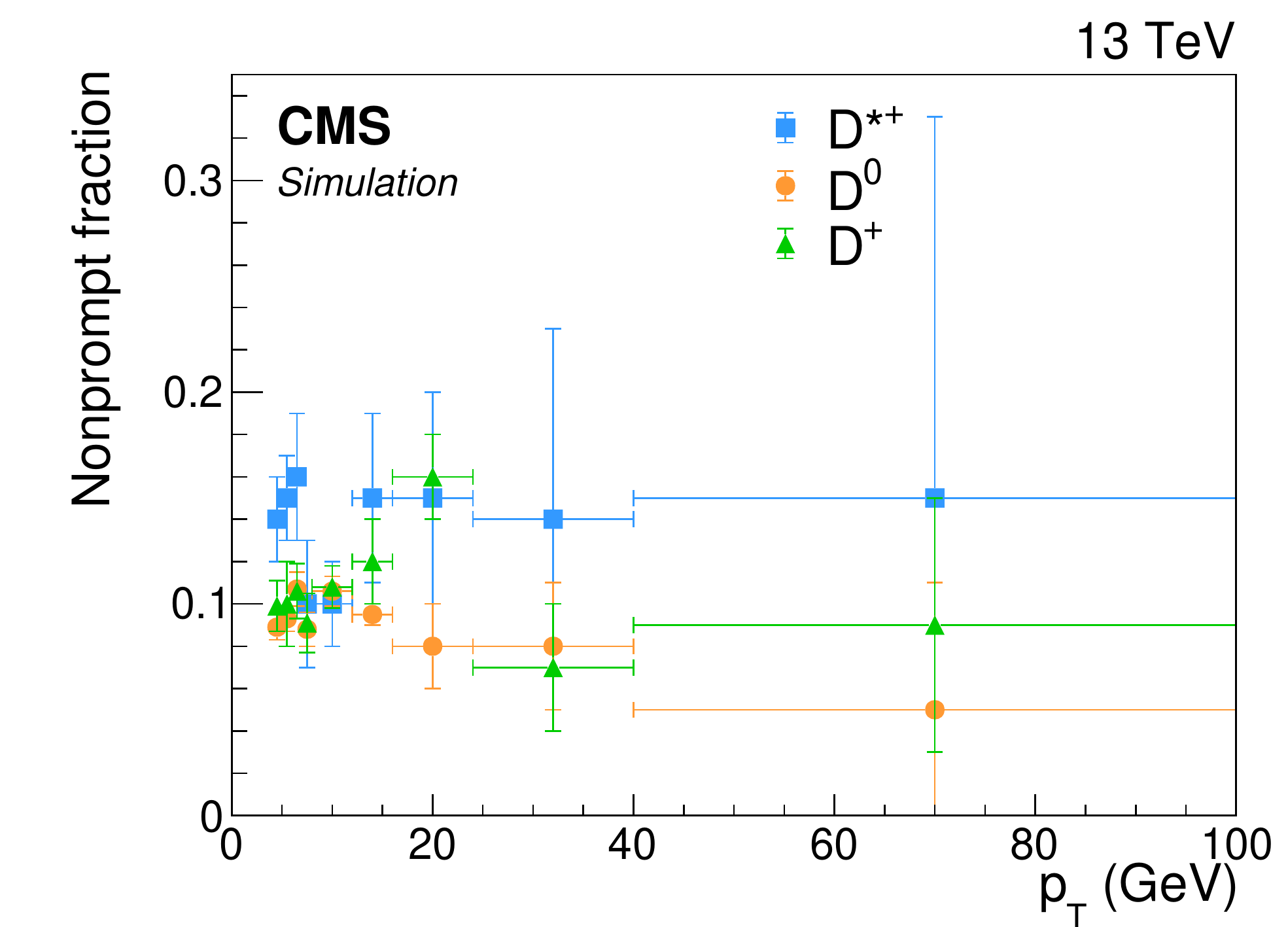}
\includegraphics[width=0.47\textwidth]{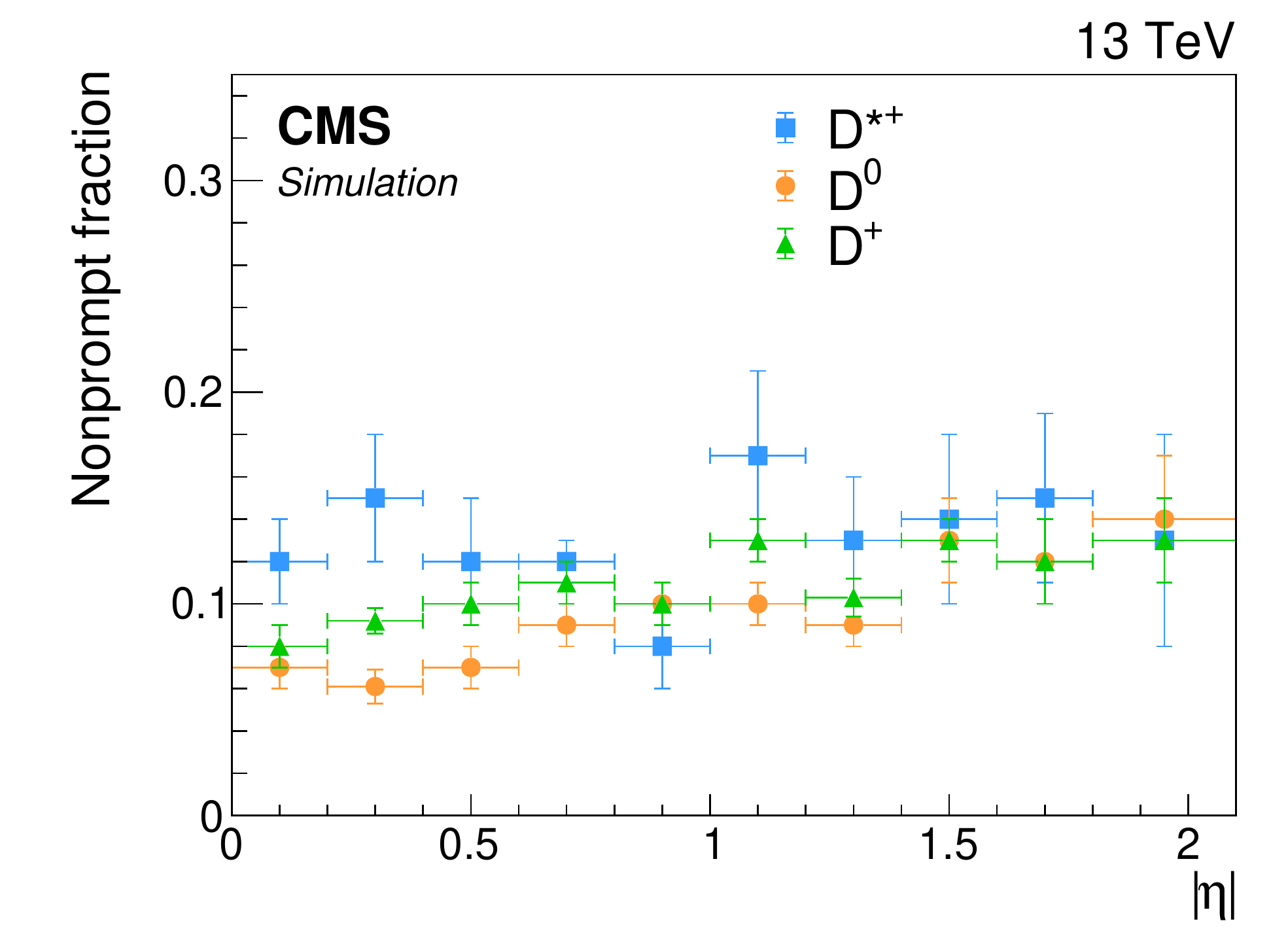}
\caption{The nonprompt fractions found from simulation, as a function of \pt (left) and $\abs{\eta}$ (right) for \PDstp (squares), \PDz (circles), and \PDp (triangles) mesons. 
The vertical lines represent the statistical uncertainties and the horizontal lines the bin widths.}
\label{contam}
\end{figure}

The aim of this work is to measure the prompt open-charm production cross sections. 
Thus, it is important to evaluate and subtract the contribution coming from nonprompt charm mesons arising from \Pb hadron decays.
Since consistency with the PV is already part of the selection requirements, the prompt signals and secondary-decay components have similar kinematic variable distributions. 
The nonprompt-background fraction is estimated from simulation, using minimum-bias events generated with the \PYTHIA8 tune CUETP8M1 \cite{CUETP}.
From the generator-level information, two subsamples are identified as being representative of the prompt and nonprompt charm meson contributions.
The same reconstruction strategy as the one described in Section \ref{reco} is applied to each of them, and the yields are computed following the method for yield evaluation reported in Section \ref{yield}, and are labeled $N_{\text{prompt}}$ and $N_{\text{nonprompt}}$, respectively.
The contamination is then evaluated as the ratio of $N_{\text{nonprompt}}$ to the sum $(N_{\text{prompt}} + N_{\text{nonprompt}})$ for each \pt and $\abs{\eta}$ bin.
The contamination is nonnegligible, ranging from 5 to 17\%, depending on the \pt and $\abs{\eta}$ bin and on the reconstructed meson.
This is expected because the requirement on the decay length significance to reject combinatorial backgrounds tends to enhance the contribution from long-lived hadrons.
In Fig.~\ref{contam}, the nonprompt-background fractions for the three mesons are shown as a function of \pt (left) and $\abs{\eta}$ (right).

The fraction of nonprompt events in the signal samples is thus taken from simulation, after ensuring a good description of data events in all the relevant quantities. 
The fractions are found to be consistent using different generator settings and with CMS measurements at 5\TeV \cite{cmsHN_hep, cmsHN_hep2}, extrapolated to the final states and kinematic region of this measurement.
The nonprompt contribution obtained with this method is subtracted from the measured values of the visible event rate for each \pt and $\abs{\eta}$ bin to get the final prompt results. 

\section{Systematic uncertainties}
\label{syst}

Several systematic uncertainty sources are considered in the measurement of the charm meson cross sections.
The dominant effects come from the uncertainties related to the tracking efficiency and the modeling of the invariant mass distributions used in the fit for both the signal and background components.
The uncertainties considered in this analysis can be organized into three different categories: decay mode and kinematic bin dependent, only decay mode dependent, and independent of both decay mode and bin. 

The first category includes the uncertainty in the efficiency coming from the finite number of MC simulation events, resulting in systematic uncertainties of 0.3, 0.3, and 3.5\%, respectively, for the \PDstp, \PDz, and \PDp meson cross sections. 
The last uncertainty is larger than the other two since the sample was enriched in $\PDstp \to \PDz\pis \to \PKm\Pgpp\pis$ decays.
The uncertainties in the nonprompt event contamination have been treated similarly, considering the limited number of events in the MC simulation. 
The uncertainties in the CMS nonprompt charm measurement \cite{cmsHN_hep2} and its extrapolation are propagated as an additional uncertainty of 2\%, applied to the three mesons.
The resulting systematic uncertainties are 3.5, 2.2, and 2.4\% for the \PDstp, \PDz, and \PDp cross sections, respectively.

The mass arbitration requirement can produce a peaking background for the small number of events affected by the arbitration, which contributes to the first uncertainty category.
This was studied in simulation by selecting a control region that does not contain events with charm mesons and was found to have a nonnegligible contribution only for the \PDp meson.
The effect for the \PDp was evaluated in each bin and found to contribute a systematic uncertainty of 8\% for $\pt < 12\GeV$, but to be negligible at higher \pt values.
This contribution is independent of $\abs{\eta}$, and an uncertainty of 6\% is assigned for all $\abs{\eta}$ bins.
This is considered as part of the background modeling systematic uncertainty.

Another source of systematic uncertainty that is included in the first category comes from the \pt selection criterion applied to the \pis in the \PDstp decay chain.
The \pt spectrum of the slow pion peaks below 0.5\GeV and the selection requirement of $\pt > 0.3\GeV$ affects the reconstruction efficiency of the \PDstp in the first \pt bin (4--5\GeV).
A systematic uncertainty of 9\% is assigned for this bin, which reflects the variation of the simulated event efficiency calculation in the two \pt bins of the \pis: $0.2 <\pt< 0.3\GeV$ and $0.3 <\pt< 0.4\GeV$. 

Since a reweighting is applied to simulated events in order to reproduce the data PU distribution, a systematic uncertainty is evaluated for each final state.
The systematic uncertainty is estimated using the weight calculated for each bin. 
The cross sections are re-evaluated using the weights raised and lowered by their statistical uncertainties.
The largest bin-by-bin change with respect to the cross section calculated with the central value of the weight is taken as the corresponding systematic uncertainty.
The total effect is 1\% for \PDstp and \PDz, and 2\% for \PDp.

The uncertainties associated with the branching fraction values and the track reconstruction efficiency depend on the decay mode but not on the candidate \pt or $\abs{\eta}$.
The first one is taken from Ref.~\cite{pdg} and has values of 1.1, 0.8, and 1.7\% for \PDstp, \PDz, and \PDp, respectively.
An uncertainty is assigned to the track reconstruction efficiency according to Ref. \cite{tracking}.
A different procedure is needed for the slow pion from the \PDstp decay.
Because of the soft \pt spectrum, a lower tracking efficiency is expected. 
The uncertainty related to the slow pion is computed by comparing the yield in data and MC simulated events when varying the \pt and $\abs{\eta}$ of the slow pion.
This results in a systematic uncertainty of 5.2\%.  
Combining the uncertainties for each track, the total systematic uncertainty from the tracking efficiency is 9.4, 4.2, and 6.1\% for the \PDstp, \PDz, and \PDp meson cross sections, respectively.

The systematic uncertainty due to the modeling of the invariant mass distribution also falls into the second category.
As described in Section \ref{yield}, the signal yields are computed by modeling the resonance peaks with a double-Gaussian function in order to take into account the different resolution effects in various kinematic regions.
The uncertainty is estimated by using instead a single-Gaussian function, the sum of three Gaussian functions, and a Crystal Ball function \cite{cb1, cb2}.
The largest deviation with respect to the default choice is then taken as the systematic uncertainty, yielding 3.6, 5.0, and 4.2\% for the \PDstp, \PDz, and \PDp meson cross sections, respectively.
For the combinatorial background description, the systematic uncertainty is evaluated by replacing the baseline function with a fourth-degree polynomial, resulting in 1.2, 4.8, and 5.3\% for the \PDstp, \PDz, and \PDp meson cross sections, respectively.

The last category, containing uncertainties independent of both decay modes and kinematic variables, includes those due to data-taking conditions and variation in detector performance.  
The systematic uncertainty in the integrated luminosity for 2016 is 2.5\% \cite{lumi}.
During the 2016 run, the CMS tracker suffered some time-dependent inefficiencies that resulted in a nonnegligible change in charm meson yields. 
The average PU rate also varied during the data taking. 
Both these effects are taken into account by correcting the different runs for the tracker inefficiency, which was determined from simulation after the PU distribution in the simulated events was reweighted to match the data.  
The resulting systematic uncertainty in the correction is estimated to be 1.4\%.

All the systematic uncertainties, except the one applied only to the first \pt bin of the \PDstp, are summarized in Table \ref{tab:syst}.
For the bin-dependent systematic uncertainties, the value given in Table \ref{tab:syst} is an average that is weighted by the number of signal events.
The total uncertainty is evaluated as the sum in quadrature of the individual contributions.

\begin{table}[ht!]
 \topcaption{Summary of the systematic uncertainties (\%) in the \PDstp, \PDz, and \PDp meson cross sections. 
For the bin-dependent systematic uncertainties in the table, the weighted average is shown. 
The total uncertainty is the sum in quadrature of the individual contributions.}
  \centering
  \begin{tabular}{lccc}
   Source & \PDstp  & \PDz  & \PDp \\
   \hline
   Signal efficiency calculation& 0.3 & 0.3 & 3.5 \\
   Nonprompt contamination      & 3.5 & 2.2 & 2.4 \\
   PU reweighting               & 1.0 & 1.0 & 2.0  \\ 
   Branching fraction           & 1.1 & 0.8 & 1.7 \\
   Tracking  efficiency         & 9.4 & 4.2 & 6.1 \\ 
   Signal modeling              & 3.6 & 5.0 & 4.2  \\
   Background modeling          & 1.2 & 4.8 & 8.0 \\
   Integrated luminosity        & 2.5 & 2.5 & 2.5 \\
   Time-dependent inefficiency  & 1.4 & 1.4 & 1.4 \\[\cmsTabSkip]  
   Total                        & 11.2& 9.0 & 12.3  \\
  \end{tabular}
  \label{tab:syst}
\end{table}

\section{Results}
\label{results}

The differential cross sections for prompt charm meson production as a function of \pt and $\abs{\eta}$ are determined using the equations:
\begin{equation}
\frac{\rd\sigma(\Pp{}\Pp \to \PD X)}{\rd\pt}= \frac{N_i(\PD \to f)}{\Delta {\pt}_{i} \; \mathcal{B}(\PD \to f)\; \mathcal{L} \; \varepsilon_{i} (\PD \to f)}\,, 
\end{equation}
\begin{equation}
\frac{\rd\sigma(\Pp{}\Pp \to \PD X)}{\rd\abs{\eta}} = \frac{N_i(\PD \to f)}{\Delta \eta_i \; \mathcal{B}(\PD \to f) \; \mathcal{L} \; \varepsilon_{i}(\PD \to f)}\,, 
\end{equation}
where $N_i(\PD \to f)$ is the number of prompt charm mesons reconstructed in the selected final state $f$ (including the charge-conjugate final state) for each bin i, after subtracting the nonprompt backgrounds, $\Delta$\pt and $\Delta \eta = 2\Delta \abs{\eta}$ are the bin widths, $\mathcal{B}(\PD \to f)$ is the branching fraction of the reconstructed decay, $\varepsilon_{i}(\PD \to f)$ is the total reconstruction efficiency of the decay chain evaluated using simulated events, and $\mathcal{L}$ is the integrated luminosity.

In Tables \ref{xsec_pt} and \ref{xsec_eta}, the differential cross section values and their uncertainties are reported for each \pt and $\abs{\eta}$ bin, respectively.
The first uncertainty is statistical and the second is systematic.
The latter is the dominant one in the majority of the \pt and $\abs{\eta}$ bins.

\begin{table}[ht!]
 \topcaption{The differential cross sections of prompt \PDstp + \PDstm, \PDz + \PaDz, and \PDp + \PDm production in \pt bins with $\abs{\eta} < 2.1$; the first uncertainty is statistical, the second is systematic.}
 \centering
 \cmsTable{ 
  \begin{tabular}{lccc}
    & \multicolumn{3}{c}{$\rd\sigma/\rd\pt$ ($\mu \Pb$/\GeVns)} \\
   \pt bin [\GeVns{}]  & \PDstp + \PDstm  & \PDz + \PaDz  & \PDp + \PDm \\
   \hline
    \begin{tabular}{r@{--}l}
      4 & 5 \\
      5 & 6 \\
      6 & 7 \\
      7 & 8 \\
      8 & 12 \\
     12 & 16 \\
     16 & 24 \\
     24 & 40 \\
     40 & 100 \\
      \end{tabular}
     &
     \begin{tabular}{r@{ $\pm$ }c@{ $\pm$ }l}
     166 & 7 & 24 \\
     96 & 6 & 11 \\ 
     47 & 3 & 5 \\
     25.6 & 1.7 & 2.9 \\ 
     8.8 & 0.2 & 1.0 \\
     1.70 & 0.09 & 0.20 \\
     (3.52 & 0.15 & 0.42) $\times 10^{-1}$ \\ 
     (4.1  & 0.3 &  0.6) $\times 10^{-2}$ \\
     (2.3  & 0.4 &  0.5) $\times 10^{-3}$ \\
      \end{tabular}
     & 
     \begin{tabular}{r@{ $\pm$ }c@{ $\pm$ }l}
     430 & 58 & 38 \\
     230 & 17 & 21 \\
     136 & 11 & 12 \\ 
     66  &  8 & 6 \\
     21.0 & 0.7 & 1.9  \\
     3.93 & 0.19 & 0.44 \\
     (8.1 & 0.5 & 0.9) $\times 10^{-1}$ \\
     (9.7 & 0.9 & 1.3) $\times 10^{-2}$ \\
     (3.3 & 0.7 & 0.8) $\times 10^{-3}$
     \end{tabular}
     &
     \begin{tabular}{r@{ $\pm$ }c@{ $\pm$ }l}
     250 & 61 & 34 \\
     79 & 11 & 11\\
     50 & 2 & 7 \\
     29.5 & 1.4 & 4.3\\
     9.4 & 0.3 & 1.2 \\
     2.05 & 0.07 & 0.24 \\
     (4.06 & 0.17 & 0.51) $\times 10^{-1}$ \\
     (6.0  & 0.6 & 1.0) $\times 10^{-2}$ \\
     (2.3  & 0.3 & 1.7) $\times 10^{-3}$ \\
     \end{tabular} 
  \end{tabular}
  }
  \label{xsec_pt}
\end{table}

\begin{table}[ht!]
 \topcaption{The differential cross sections of prompt \PDstp + \PDstm, \PDz + \PaDz, and \PDp + \PDm production in $\abs{\eta}$ bins with $4 < \pt < 100\GeV$; the first uncertainty is statistical, the second is systematic.}
  \centering
  \begin{tabular}{lccc}
      & \multicolumn{3}{c}{ $\rd\sigma/\rd\abs{\eta}$ ($\mu \Pb$)}\\  
   $\abs{\eta}$ range & \PDstp + \PDstm & \PDz + \PaDz & \PDp + \PDm  \\
   \hline
    \begin{tabular}{r@{--}l}
      0 & 0.2 \\
      0.2 & 0.4 \\
      0.4 & 0.6 \\
      0.6 & 0.8 \\
      0.8 & 1.0 \\
      1.0 & 1.2 \\
      1.2 & 1.4 \\
      1.4 & 1.6 \\
      1.6 & 1.8 \\
      1.8 & 2.1 \\
     \end{tabular}
     &
     \begin{tabular}{r@{ $\pm$ }c@{ $\pm$ }l}
     92 &  3 & 10  \\
     96 &  4 & 11 \\
     94 &  3 & 11 \\ 
     96 &  4 & 11 \\
     111 & 6 & 12 \\
     97 & 8 & 11 \\
     106 & 8 & 12 \\
     81 & 8 &  9 \\
     85 & 13 & 10 \\ 
     70 &  9 & 17 \\
     \end{tabular} 
     &
     \begin{tabular}{r@{ $\pm$ }c@{ $\pm$ }l}
     210 & 7 & 19 \\
     240 &  8 & 21 \\
     243 &  9 & 22 \\
     255 & 10 & 23 \\
     220 & 13 & 20 \\
     265 & 20 & 24 \\
     250 & 24 & 23 \\
     265 & 27 & 24 \\
     298 & 45 & 27 \\
     291 & 45 & 64 \\  
     \end{tabular}  
     &
     \begin{tabular}{r@{ $\pm$ }c@{ $\pm$ }l}
     87 & 3 & 11 \\
     90 & 3 & 12 \\ 
     86 & 6 & 11 \\
     93 & 7 & 12 \\
     102 & 5 & 19  \\
     107 & 5 & 20 \\
     106 & 6 & 29 \\
     109 & 12 & 28  \\
     112 & 11 & 24 \\
     123 & 22 & 27 \\ 
     \end{tabular}  
  \end{tabular}
  \label{xsec_eta}
\end{table}

\begin{figure}[!hp]
\centering
\includegraphics[width=0.65\textwidth]{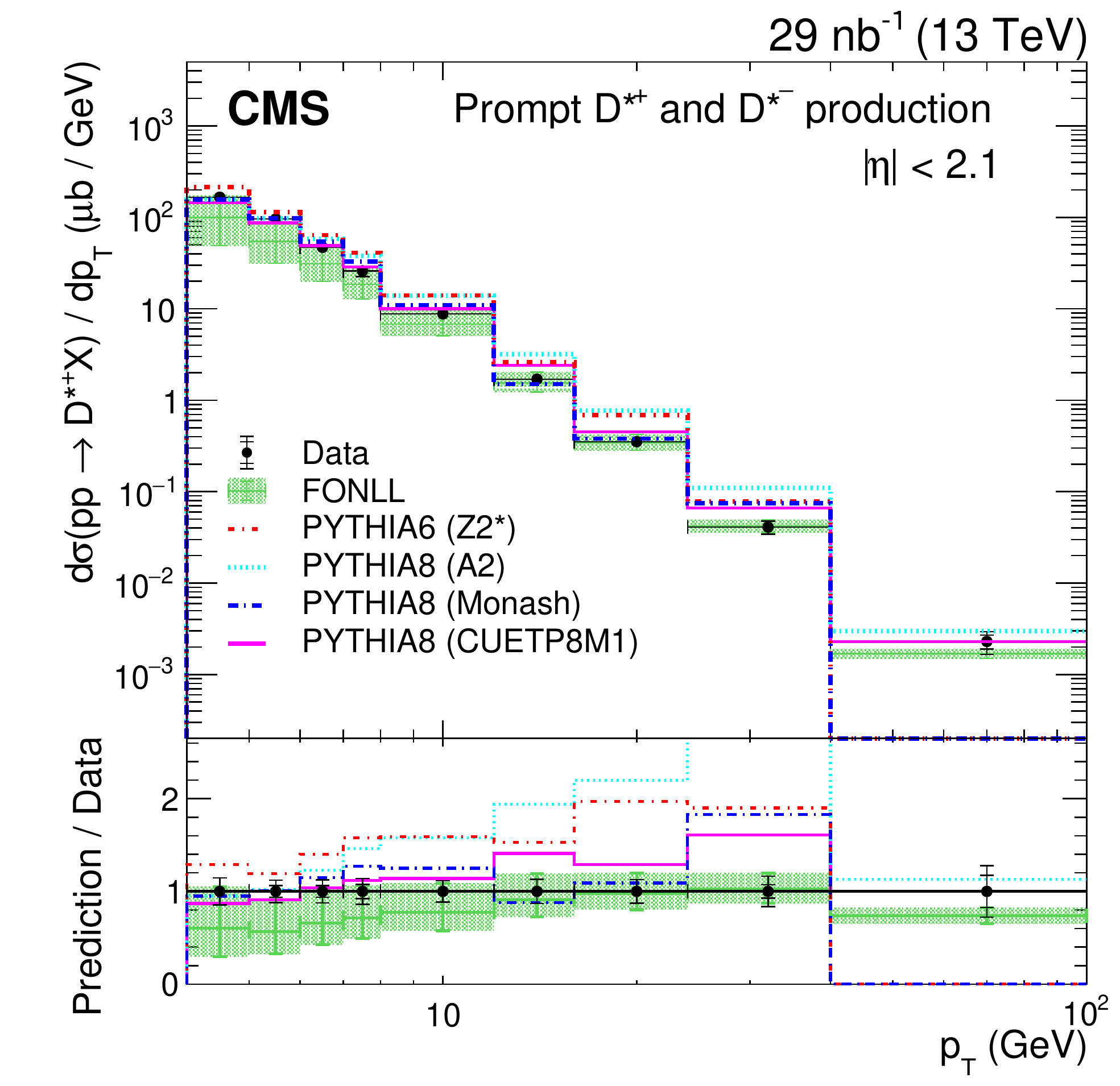}
\includegraphics[width=0.65\textwidth]{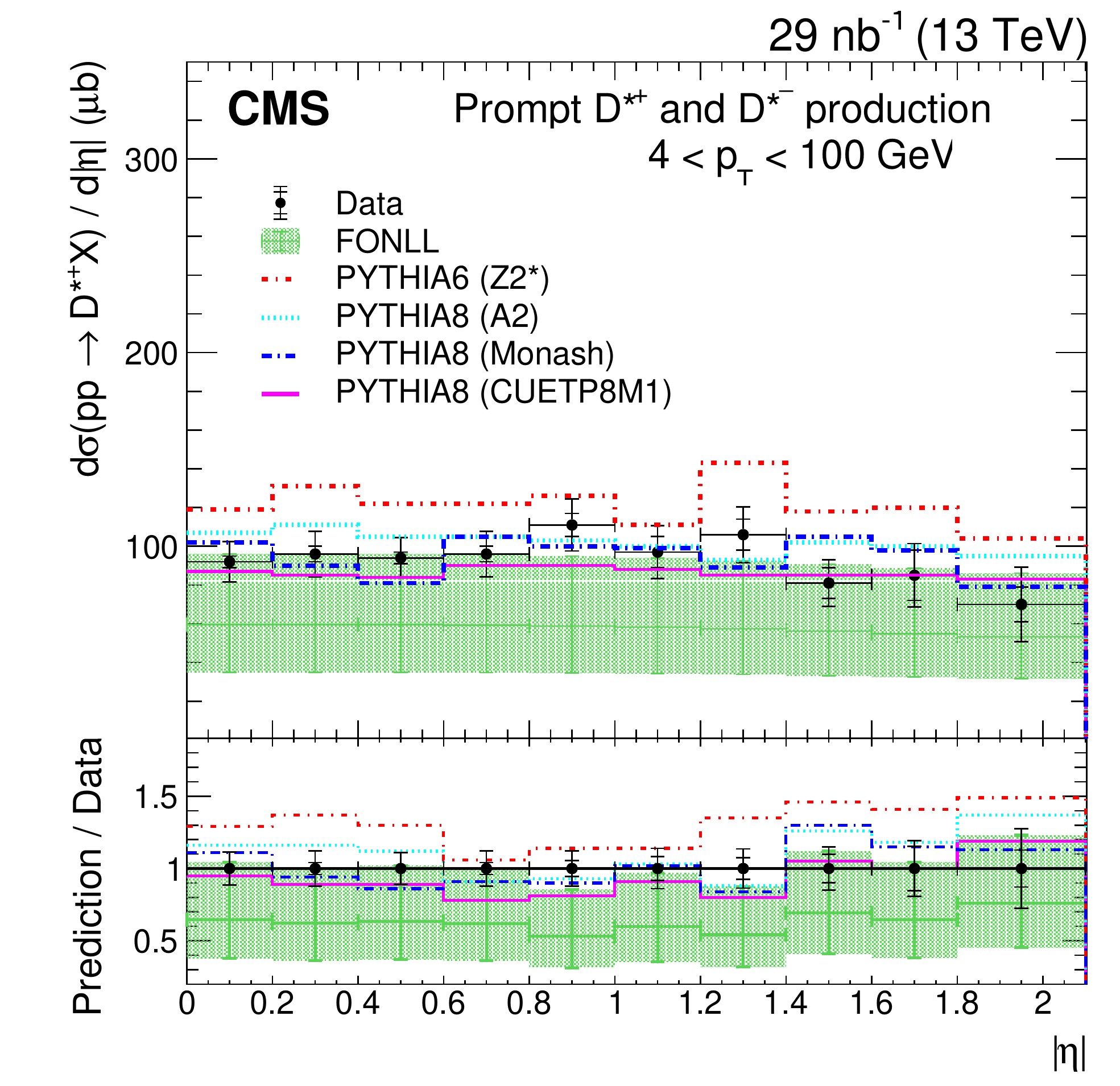}
\caption{Differential cross sections $\rd\sigma/\rd\pt$ (upper) and $\rd\sigma/\rd\abs{\eta}$ (lower) for prompt \PDstpm meson production. 
Black markers represent the data and are compared with several MC simulation models and theoretical predictions. 
The statistical and total uncertainties are shown by the inner and outer vertical lines, respectively.
The FONLL band represents the standard uncertainties in the prediction as detailed in the text. 
The lower panel gives the ratios of the predictions to the data.}
\label{DS_xsec}
\end{figure}

\begin{figure}[!hp]
\centering
\includegraphics[width=0.65\textwidth]{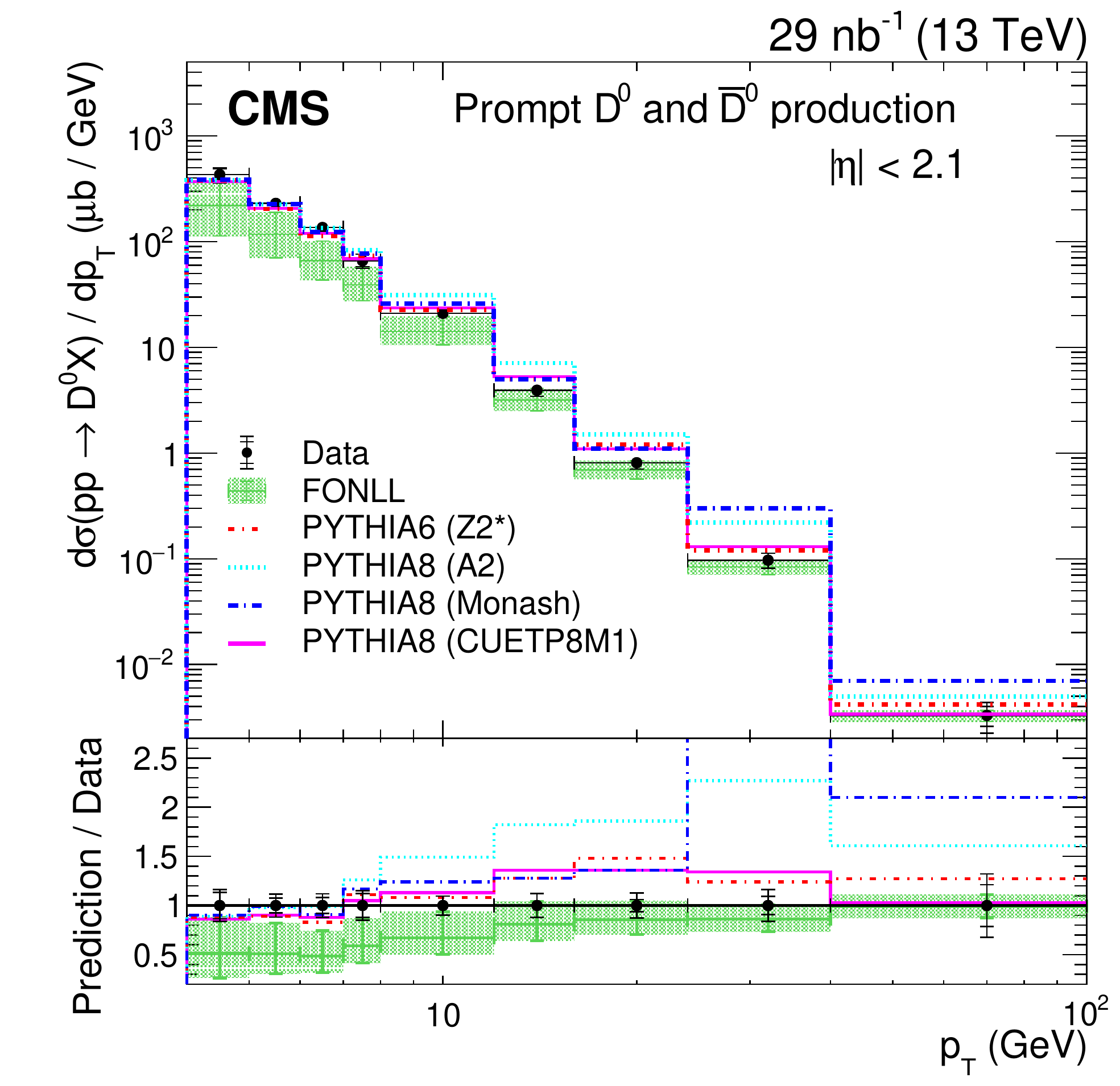}
\includegraphics[width=0.65\textwidth]{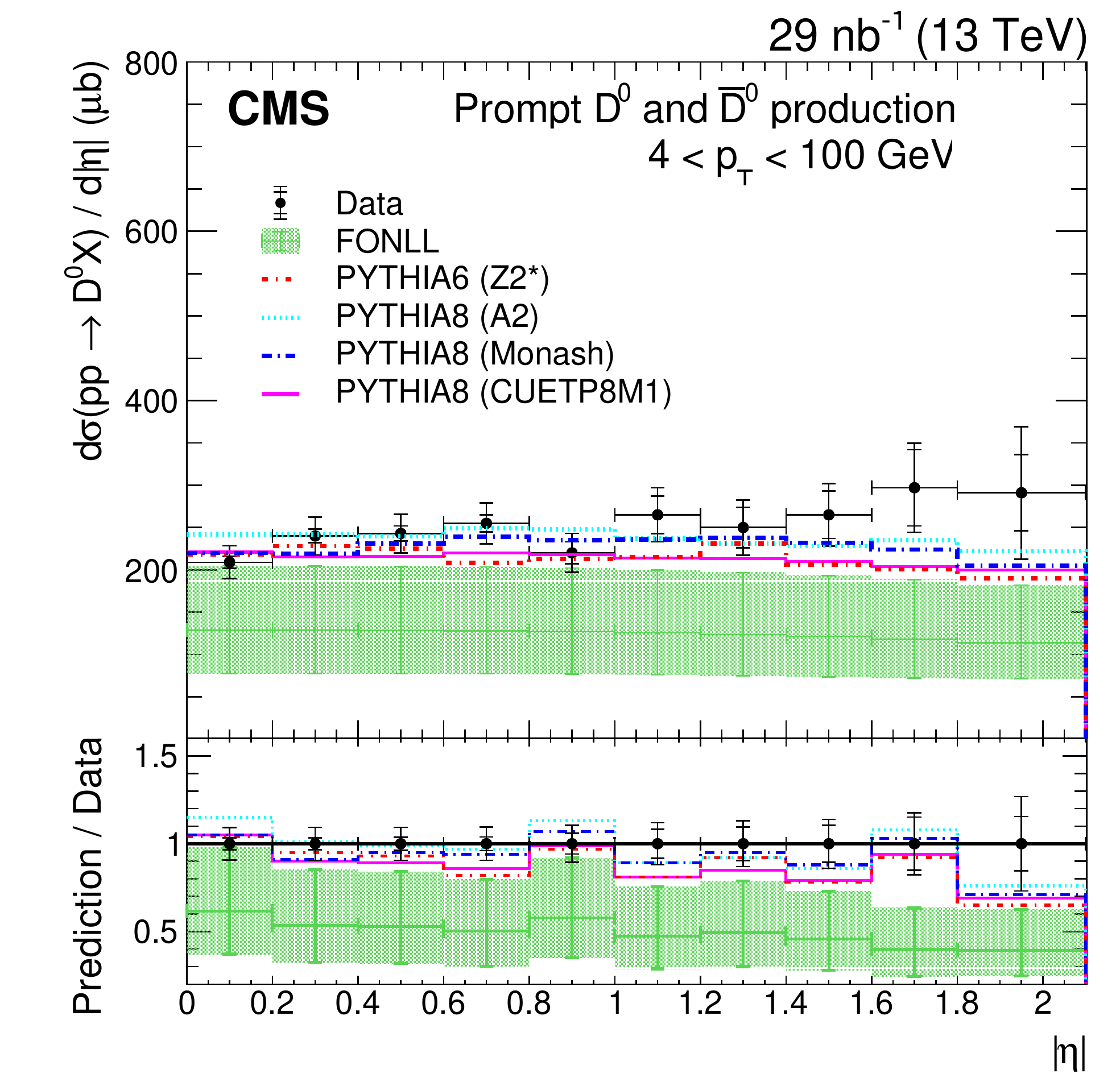}
\caption{Differential cross section $\rd\sigma/\rd\pt$ (upper) and $\rd\sigma/\rd\abs{\eta}$ (lower) for prompt \PDz(\PaDz) meson production. 
Black markers represent the data and are compared with several MC simulation models and theoretical predictions. 
The statistical and total uncertainties are shown by the inner and outer vertical lines, respectively.      The FONLL band represents the standard uncertainties in the prediction as detailed in the text. 
The lower panel gives the ratios of the predictions to the data.}
\label{D0_xsec}
\end{figure}

\begin{figure}[!hp]
\centering
\includegraphics[width=0.65\textwidth]{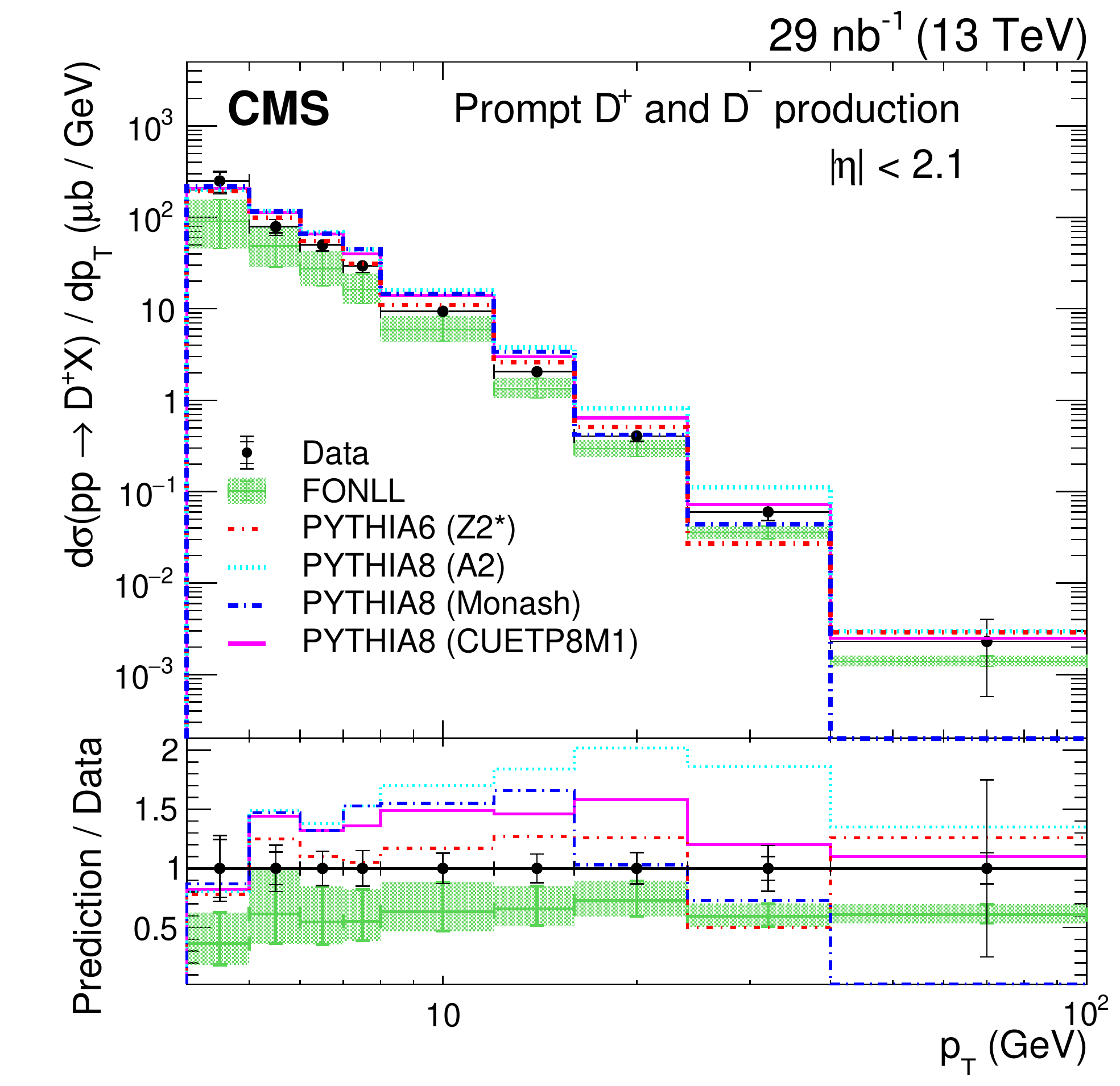}
\includegraphics[width=0.65\textwidth]{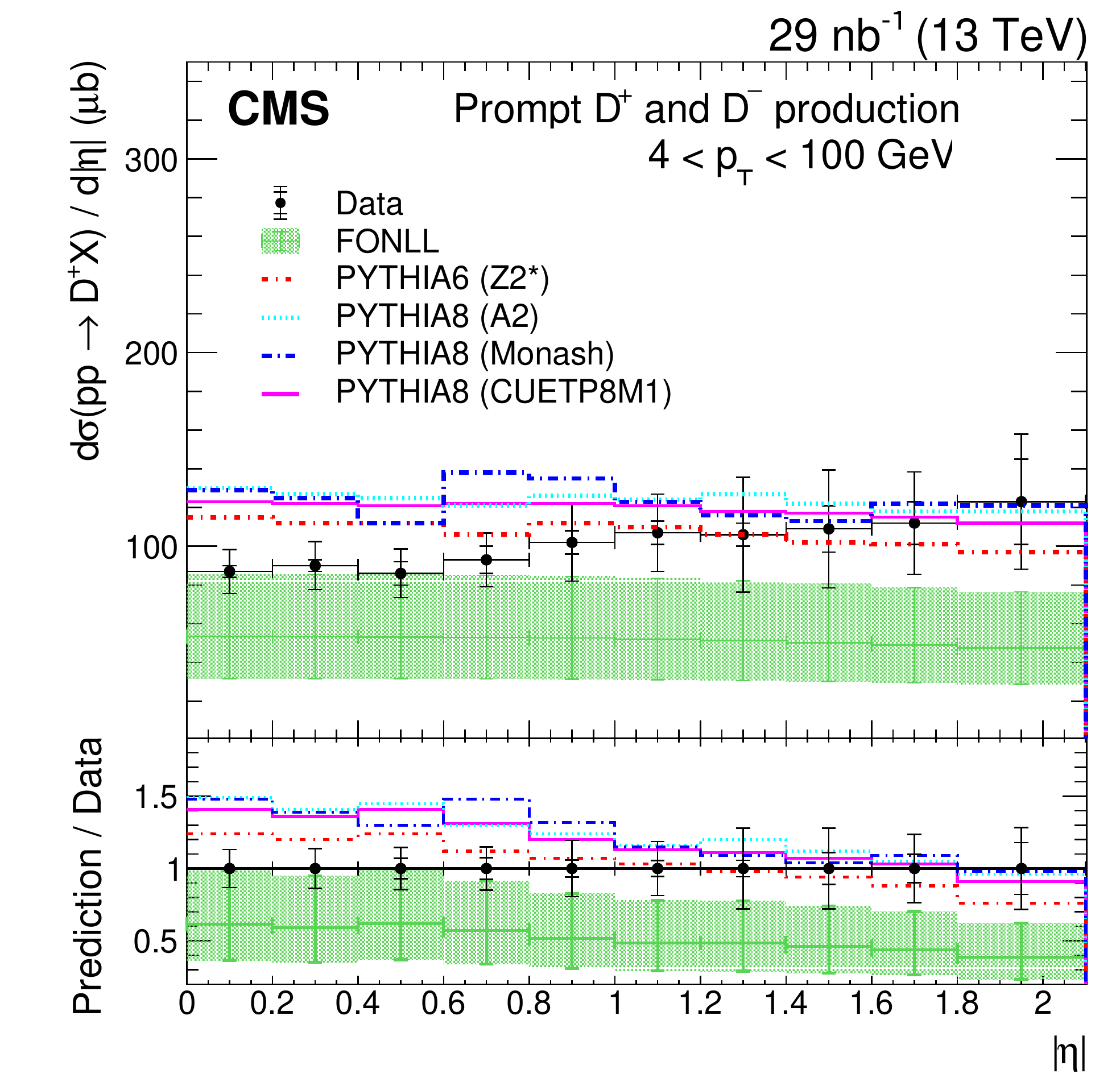}
\caption{Differential cross section $\rd\sigma/\rd\pt$ (upper) and $\rd\sigma/\rd\abs{\eta}$ (lower) for prompt \PDpm meson production. 
Black markers represent the data and are compared with several MC simulation models and theoretical predictions. 
The statistical and total uncertainties are shown by the inner and outer vertical lines, respectively.      The FONLL band represents the standard uncertainties in the prediction as detailed in the text.
The lower panel gives the ratios of the predictions to the data.}
\label{D_xsec}
\end{figure}

The differential cross sections for the three charm mesons, as a function of \pt (upper) and $\abs{\eta}$ (lower), are shown in Figs. \ref{DS_xsec}--\ref{D_xsec}, where the data points are compared to several MC predictions and theoretical calculations.
The cross section values are compared to:

\begin{itemize}
\item the predictions at next-to-leading-order (NLO) plus next-to-leading-logarithmic accuracy from FONLL \cite{fonll, fonll2} calculations, shown as bands representing the upper and lower limits for a given \pt bin, as detailed below. 
The parton distribution function (PDF) CTEQ6.6 has been used. 
The renormalization ($\mu_\mathrm{R}$) and factorization ($\mu_\mathrm{F}$) scales are set to $\mu_\mathrm{R} = \mu_\mathrm{F} = \mu_0$ where $\mu_0$ is defined as $\mu_0^2=m^2_\PQc+{\pt}^2_{\PQc}$ and $m_\PQc = 1.5 \GeV$ and ${\pt}_{\PQc}$ are the mass and transverse momentum of the charm quark, respectively. 
No default fragmentation fractions are provided for the heavy-hadron fragmentation; the fragmentation fractions used are 0.243, 0.609, and 0.240 for the \PDstp, \PDz, and \PDp, respectively \cite{ff}. 
The parameter variations used for the evaluation of the uncertainty bands are (all variations are added in quadrature): \begin{itemize}
\item scale uncertainties: $\mu_0/2 < \mu_\mathrm{R}$, $\mu_\mathrm{F} < 2 \mu_0$ with $1/2 < \mu_\mathrm{R}/\mu_\mathrm{F} < 2$;
\item charm mass variation: $m_\PQc = 1.3$--1.7\GeV;
\item PDF uncertainties: calculated according to the individual PDF set recipe.
\end{itemize}

Here the universality of charm fragmentation is implicitly still assumed, although recent results in a different kinematic range seem to demonstrate that universality is violated in a \pt-dependent way \cite{ff_un1, ff_un2, ff_un3, ff_un4, ff_un5}.
This might in principle require an evaluation of \pt-dependent downward corrections to the predicted D meson yields of order 5-20$\%$ in the \pt range measured here, while the measurements are not affected.
Since a scheme to consistently evaluate and apply such corrections to the FONLL predictions does not yet exist, and the difference is still expected to be subdominant in particular compared to the large QCD scale uncertainties, no uncertainties were assigned for this source here.

\item leading-order (LO) plus parton shower (PS) simulations using \PYTHIA 6.4 \cite{pythia6} with the tune Z2* \cite{Z2star}, which is based on the CMS Z1 tune \cite{Field:2010bc}, but adopting the CTEQ6L PDF set instead of the previous CTEQ5L. 
The tune is the result of optimizing two parameters that refer to the regularization scale: ${\pt}_{\perp 0}$, for multiple interactions at a reference energy and the power exponent of the energy rescaling used to determine the value of ${\pt}_{\perp 0}$ as it goes to zero at scales different from the reference scale;
\item LO plus PS simulations by \PYTHIA 8.202 \cite{pythia8} with the tunes: \begin{itemize}
  \item A2, which is an ATLAS minimum-bias tune \cite{A2} validated using their kinematic distributions and based on the tune 4C \cite{4C}, using the MSTW PDF;
  \item Monash \cite{monash}, which was developed by re-evaluating the constraints imposed by LEP and SLD on hadronization, in particular with regard to heavy-quark fragmentation and strangeness production; it is a \PYTHIA8 tune using the NNPDF2.3 LO PDF. 
  \item CUETP8M1 \cite{CUETP}, which is a CMS-specific tune and stands for ``CMS Underlying Event Tune \PYTHIA8". It is based on Monash (M1), but the two multiple-parton-interaction (MPI), energy-dependent parameters, which are the MPI cutoff value and the exponent $\epsilon$ of the $\sqrt{s}$ dependence, are determined by fitting underlying events in CMS data at $\sqrt{s} =$ 0.9, 1.96, and 7\TeV. 
In Ref. \cite{pythia_tunes} it was shown that neither the Monash nor the CUETP8M1 tunes describe well the central value of underlying events in data at $\sqrt{s}= 13\TeV$. 
This suggests that the tune CUETP8M1 does not produce enough charged particles at 13\TeV.
  \end{itemize}
\end{itemize}

The \PYTHIA predictions are from samples of various sizes with corresponding varying statistical uncertainties, which are not shown in the figures.
The lower panels in Figs. \ref{DS_xsec}--\ref{D_xsec} display the ratio of the FONLL and \PYTHIA predictions to the data, for which the statistical and total uncertainties are shown by the inner and outer bars, respectively.

The agreement with the different predictions is fair in the wide kinematic range analyzed.
However, none of the MC event generators models the data well over the entire measurement region.
The measurements tend to favor a higher cross section than predicted by FONLL and a lower one than estimated by \PYTHIA, although the predictions from the \PYTHIA generator are sensitive to the different tunes used.  
The cross section predictions from the different PYTHIA tunes differ in both normalization and shape, which confirms that  the description of the data provided by the models is sensitive to further model improvements.
Overall, the best description of the data is given by the upper edge of the FONLL uncertainty band.

\subsection{Comparison with other experiments}

While there are several previous LHC measurements of charm meson production cross sections, none of them cover the same kinematic region at the same center-of-mass energy as the results presented here. 
However, since the previous measurements can also be compared to FONLL predictions, it is useful to see if a consistent picture emerges.   
In the comparisons that follow, no scaling is performed on the previous measurements to account for different kinematic regions, center-of-mass energies, or cross section definitions, but the FONLL predictions were adjusted to match the conditions of the various results.

\begin{figure}[ht!]
\centering
\includegraphics[width=0.48\textwidth]{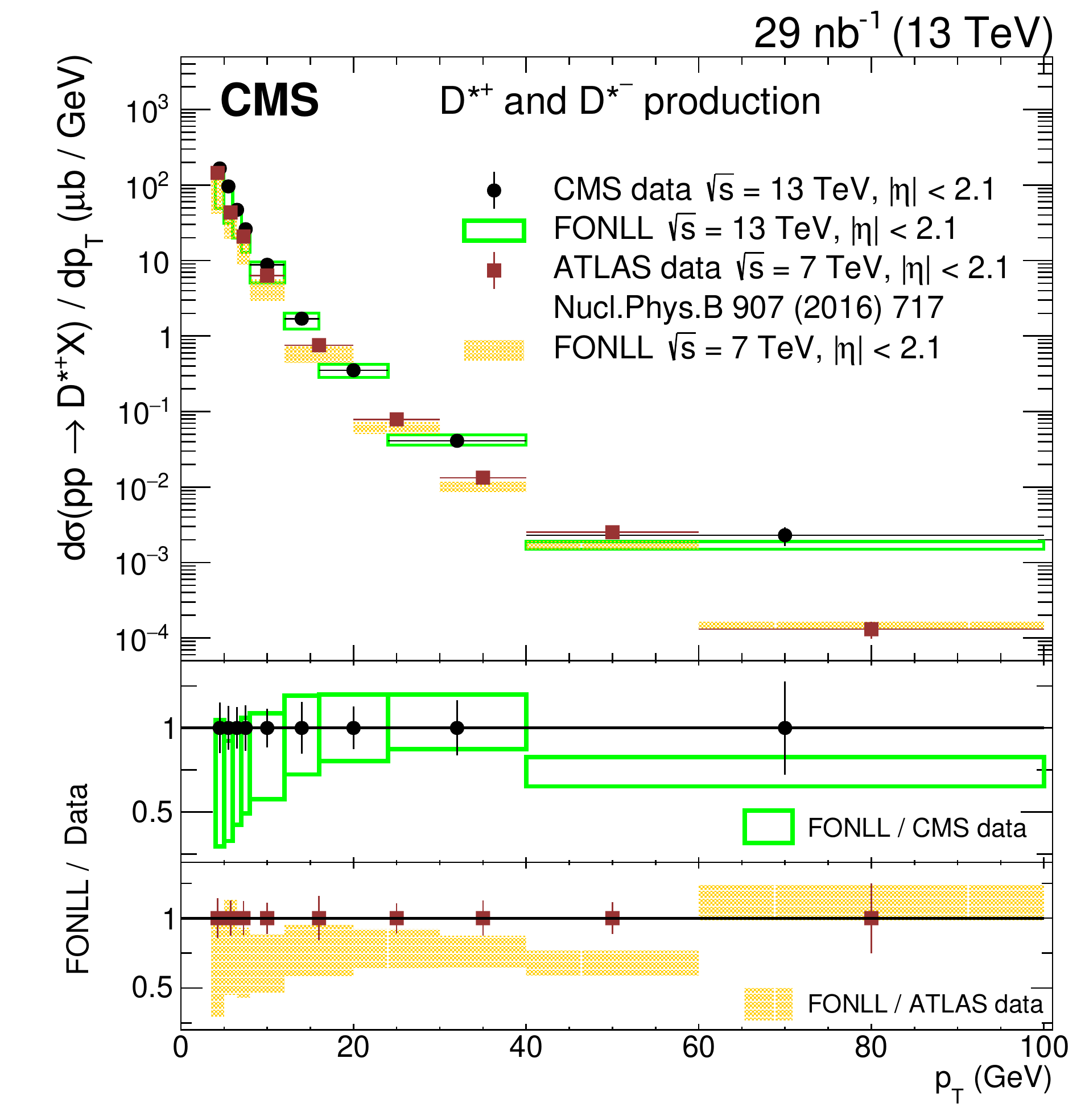}
\includegraphics[width=0.48\textwidth]{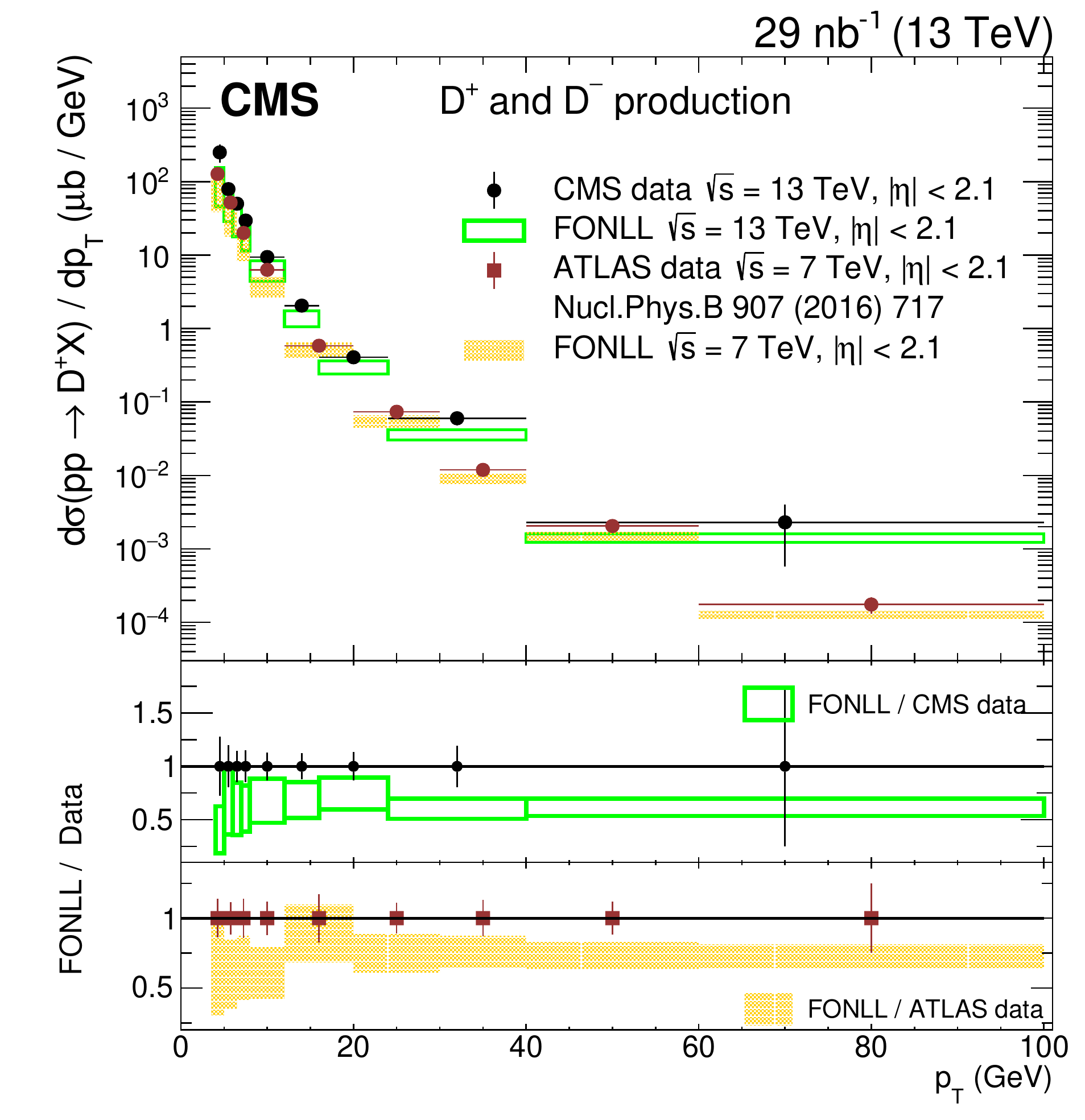}
\caption{Differential cross section $\rd\sigma/\rd\pt$ for \PDstpm (left) and \PDpm (right) meson production, comparing the production from CMS (black circles, prompt, this paper) at $\sqrt{s}= 13\TeV$ and ATLAS (red squares, prompt $+$ nonprompt) at $\sqrt{s}= 7\TeV$ \protect\cite{atlas}. 
The corresponding predictions from FONLL are shown by the unfilled and filled boxes, respectively. 
The vertical lines on the points give the total uncertainties in the data, and the horizontal lines show the bin widths.
The two lower panels in each plot give the ratios of the FONLL predictions to the CMS and ATLAS data, shown by circles and squares, respectively.}
\label{ATLAS_CMS}
\end{figure}

The measurements from the ATLAS experiment \cite{atlas}, although performed with data collected at $\sqrt{s}= 7\TeV$, are the closest to the ones presented in this paper in terms of the acceptance and kinematic regime.
Figure \ref{ATLAS_CMS} shows both the ATLAS and CMS results, compared to the respective FONLL predictions at $\sqrt{s}= 7$ and 13\TeV for both \PDstp (left) and \PDpm (right) mesons.
Since the ATLAS measurement includes both prompt and nonprompt charm mesons, the corresponding FONLL predictions include both components as well.
The lower two panels in the figure give the ratio of the FONLL predictions to the respective measurements.
The results show good agreement in terms of shape, and the comparison between the data and the theoretical predictions is very similar for the two experiments.
The central value of the FONLL predictions tends to underestimate the data, but the upper edge of the FONLL band agrees reasonably well. 

\begin{figure}[ht!]
\centering
\includegraphics[width=0.48\textwidth]{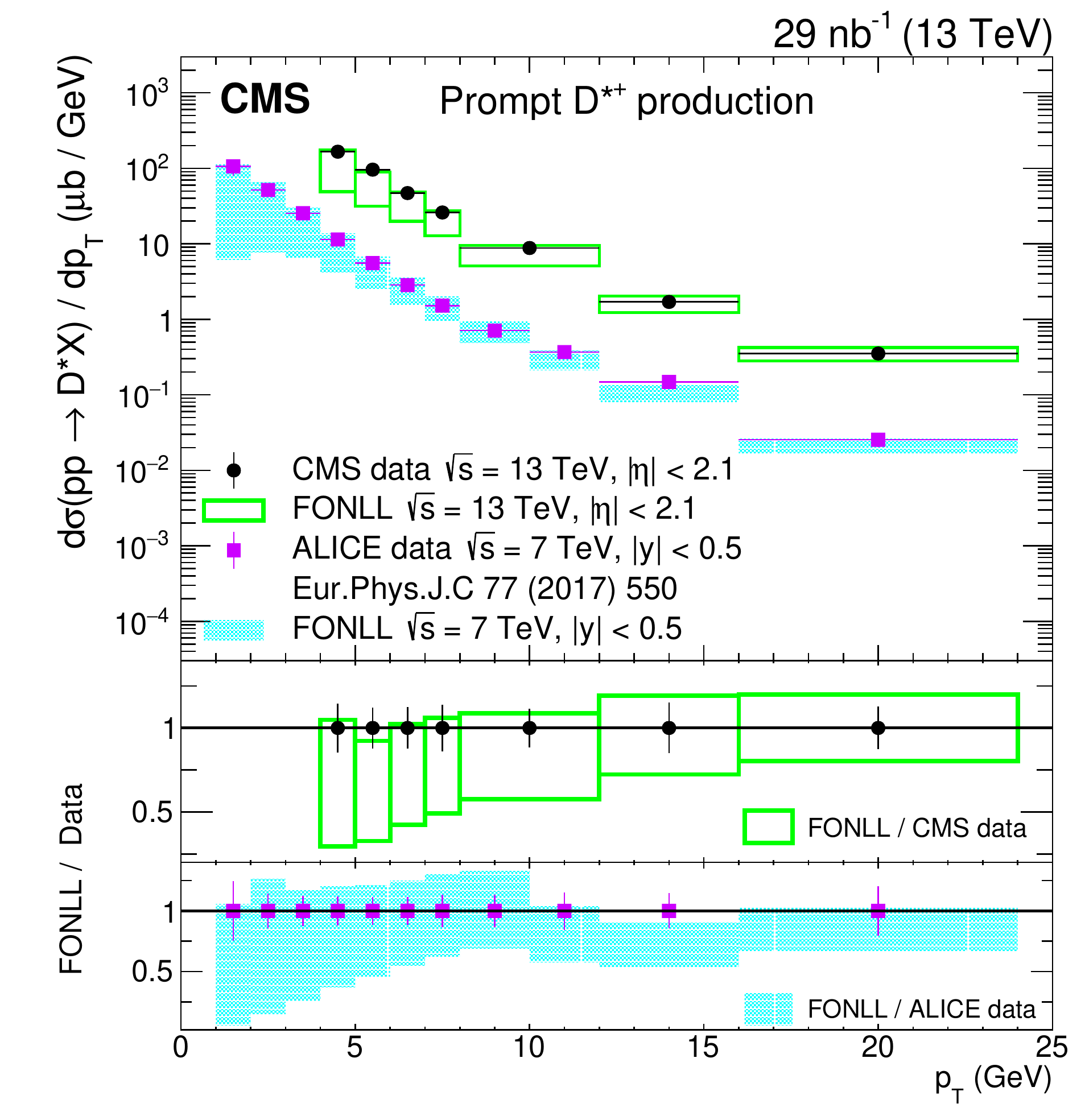}
\includegraphics[width=0.48\textwidth]{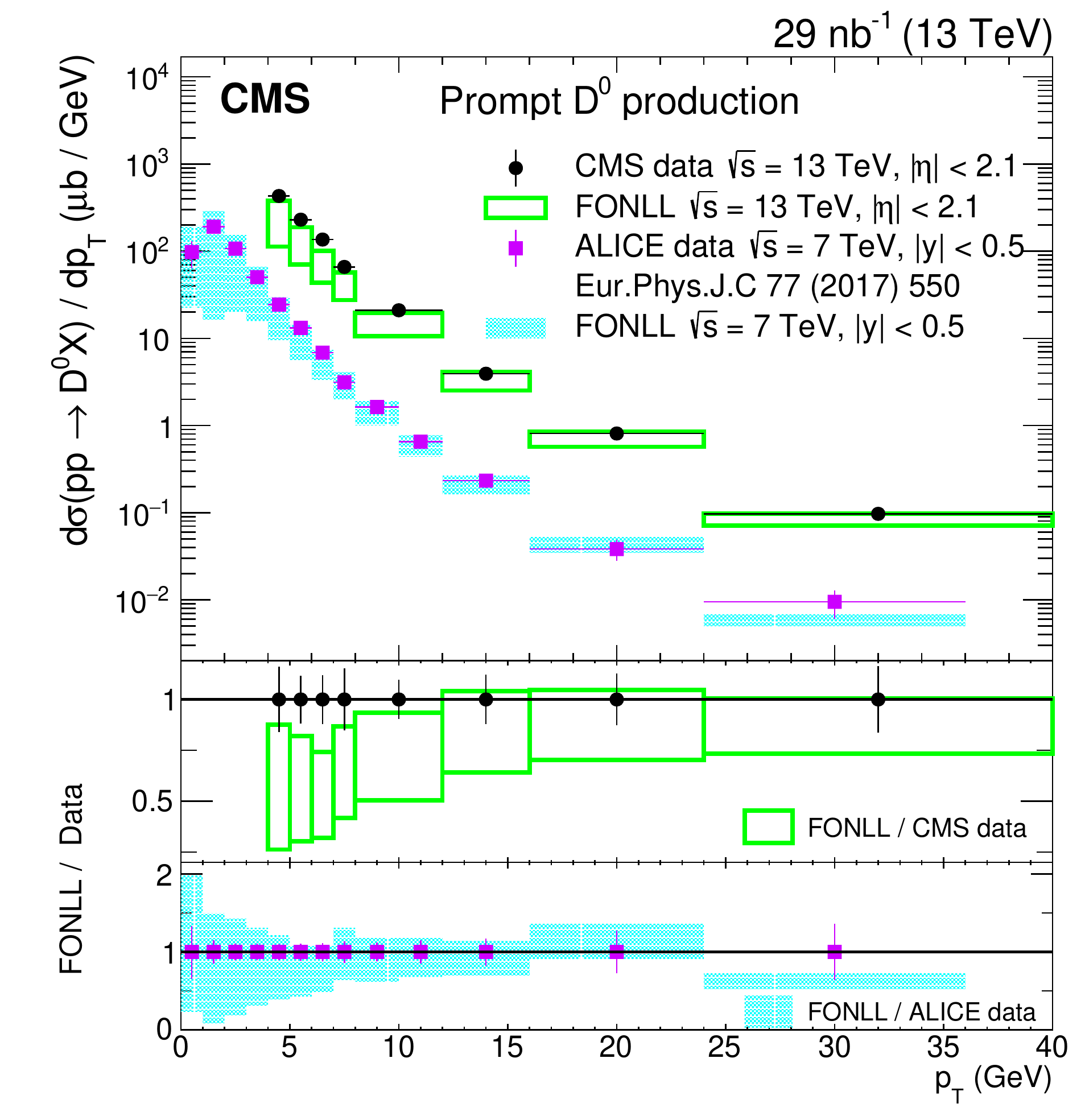}\\
\includegraphics[width=0.48\textwidth]{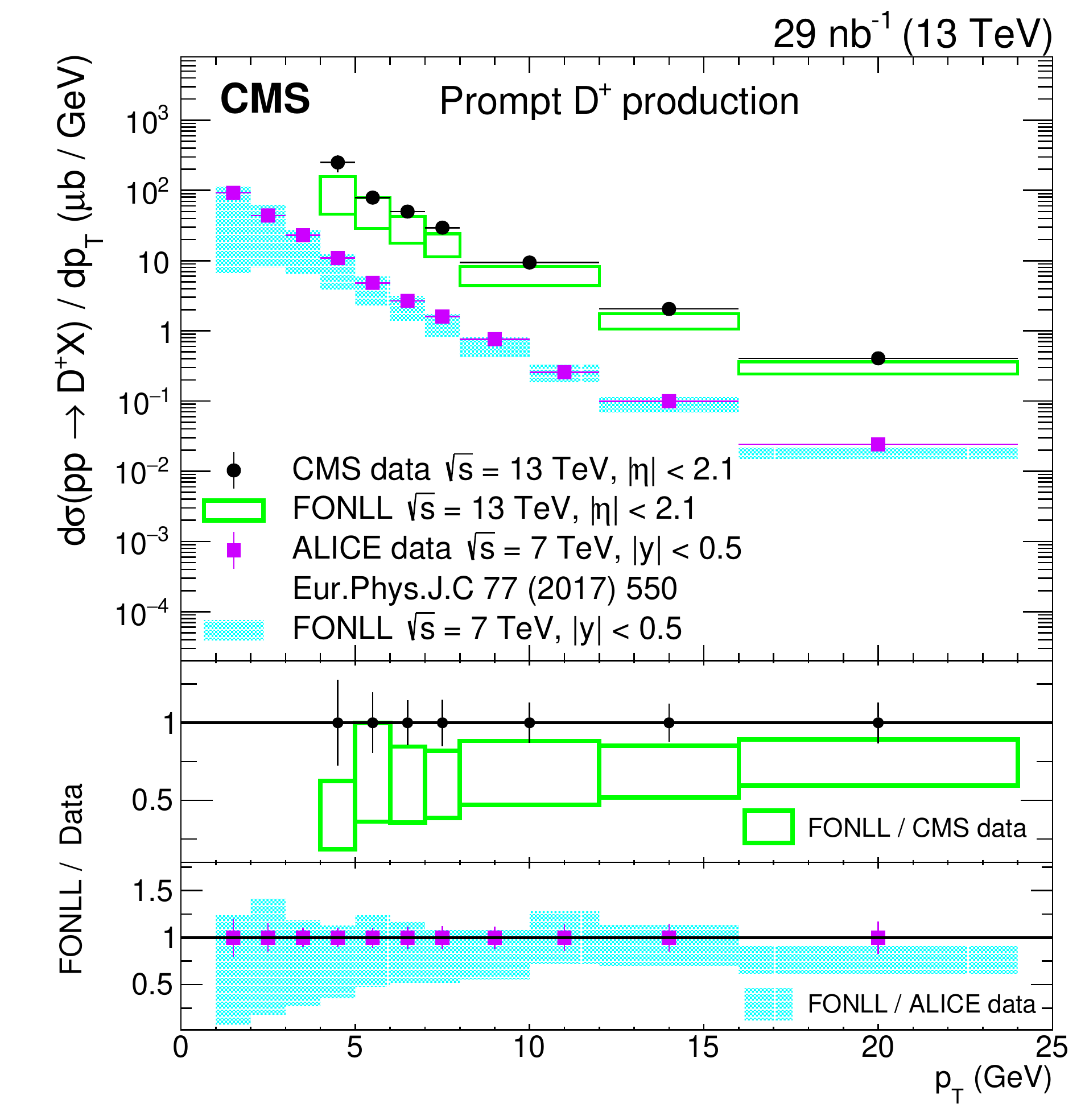}
\caption{Differential cross section $\rd\sigma/\rd\pt$ for prompt \PDstpm (upper left), \PDz$+$\PaDz (upper right) and \PDpm (lower) meson production with $\pt < 24\GeV$ from CMS (black circles, this paper) at $\sqrt{s}= 13\TeV$ and ALICE \protect\cite{alice2} (magenta squares) at $\sqrt{s}= 7\TeV$ and $\abs{y} < 0.5$. 
The corresponding predictions from FONLL are shown by the unfilled and filled boxes, respectively. 
The cross section definition by ALICE includes a factor of $1/2$ that accounts for the fact that the measured yields include particles and antiparticles while the cross sections are given for particles only. 
The same is true for the corresponding FONLL predictions, as well.
The vertical lines on the points give the total uncertainties in the data, and the horizontal lines show the bin widths.
The two lower panels in each plot give the ratios of the FONLL predictions to the CMS and ALICE data, shown by circles and squares, respectively.
}
\label{ALICE_CMS}
\end{figure}

Figure \ref{ALICE_CMS} shows the comparison with the ALICE results \cite{alice, alice2} for the \PDstp, \PDz, and \PDp cross sections at $\sqrt{s}= 7\TeV$ in the range $1 < \pt < 24\GeV$ ($0 < \pt < 36\GeV$ for the \PDz) and for the rapidity region $\abs{y} < 0.5$.
Between the two ALICE measurements, the more recent one \cite{alice2} is chosen for the comparison.
It should be noted that the cross section definition by ALICE includes a factor of $1/2$ that accounts for the fact that the measured yields include particles and antiparticles, while the cross sections are given for particles only.
The same is true for the corresponding FONLL predictions, as well.
To provide a relevant comparison, the CMS measurements are given for $\pt < 24\GeV$ ($\pt < 40\GeV$ for the \PDz), for a better comparison with the ALICE points.
Both sets of results are consistent with the respective FONLL predictions and close to their upper edge, as shown in the lower two panels. 

\begin{figure}[ht!]
\centering
\includegraphics[width=0.48\textwidth]{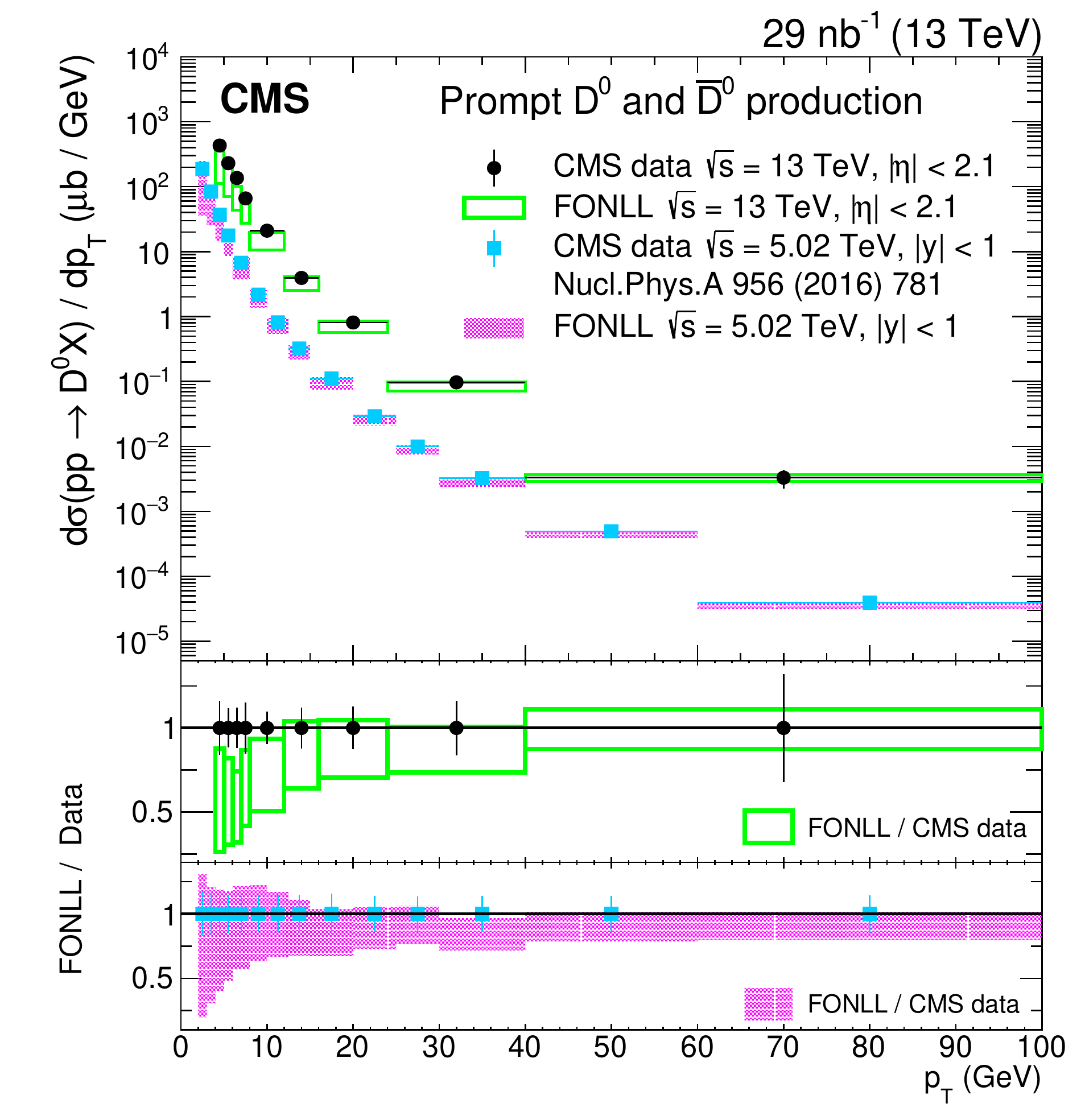}
\caption{Differential cross section $\rd\sigma/\rd\pt$ for the prompt $\PDz+\PaDz$ meson production from CMS at $\sqrt{s}= 13\TeV$ (black circles, this paper) and 5.02\TeV \protect\cite{cmsHN_hep} (light blue squares) for $\abs{y} < 1$. 
The corresponding FONLL predictions are shown by the unfilled and filled boxes, respectively. The vertical lines on the points give the total uncertainty in the data, and the horizontal lines show the bin widths.
The two lower panels give the ratios of the FONLL predictions to the CMS data at $\sqrt{s}= 13\TeV$ and 5.02\TeV, shown by circles and squares, respectively.}
\label{CMS_CMS}
\end{figure}

The CMS experiment has also measured the \PDz cross section in \Pp{}\Pp collisions at $\sqrt{s}=5.02\TeV$ for $\abs{y} < 1$ \cite{cmsHN_hep}.
Figure \ref{CMS_CMS} shows this measurements in comparison with the FONLL predictions and the corresponding 13\TeV results from this analysis.
The rise in cross section between 5.02 and 13\TeV is obvious from the figure, as well as a similar agreement between the 5.02\TeV measurements and the FONLL predictions.

\begin{figure}[ht!]
\centering
\includegraphics[width=0.48\textwidth]{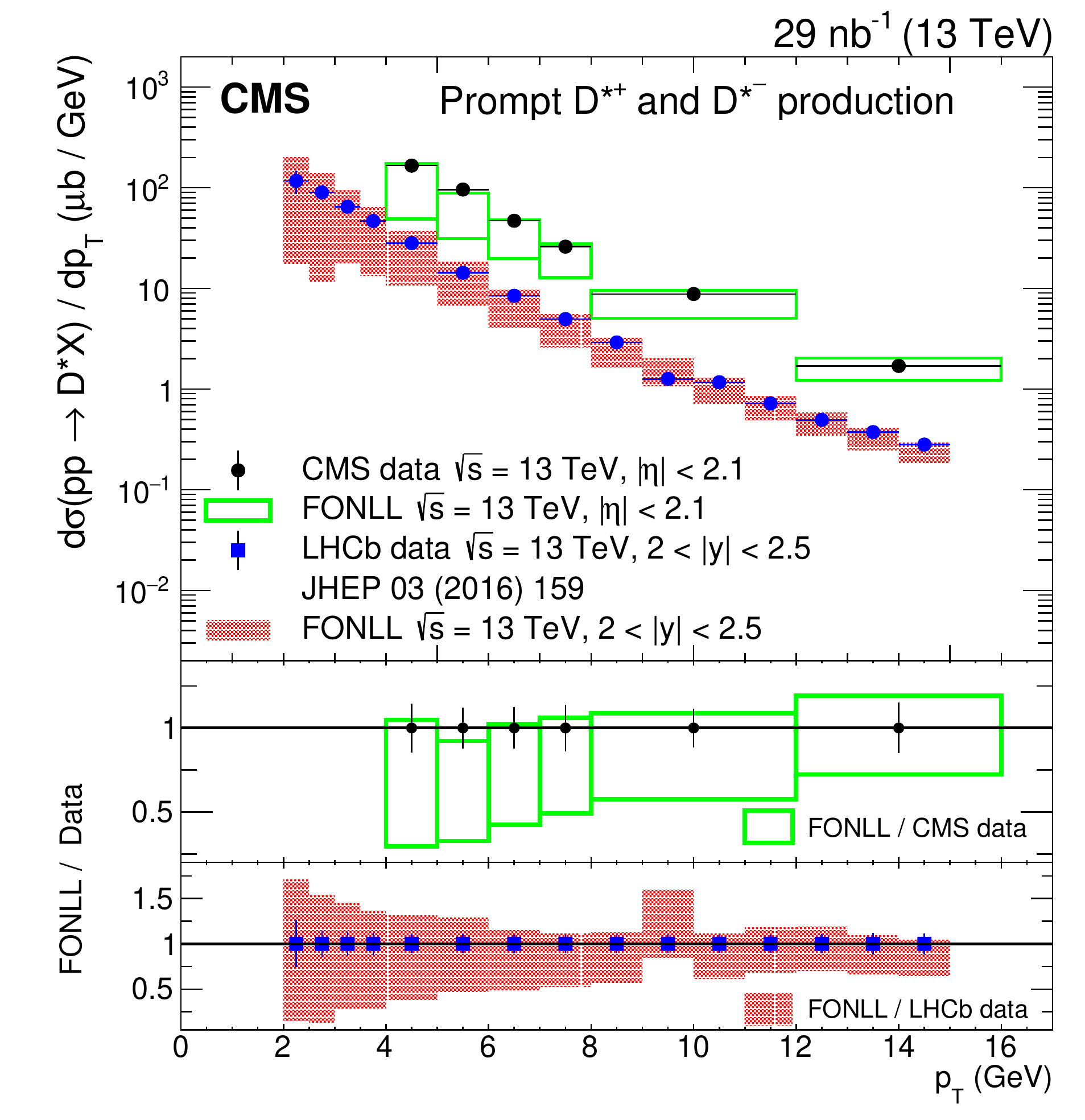}
\includegraphics[width=0.48\textwidth]{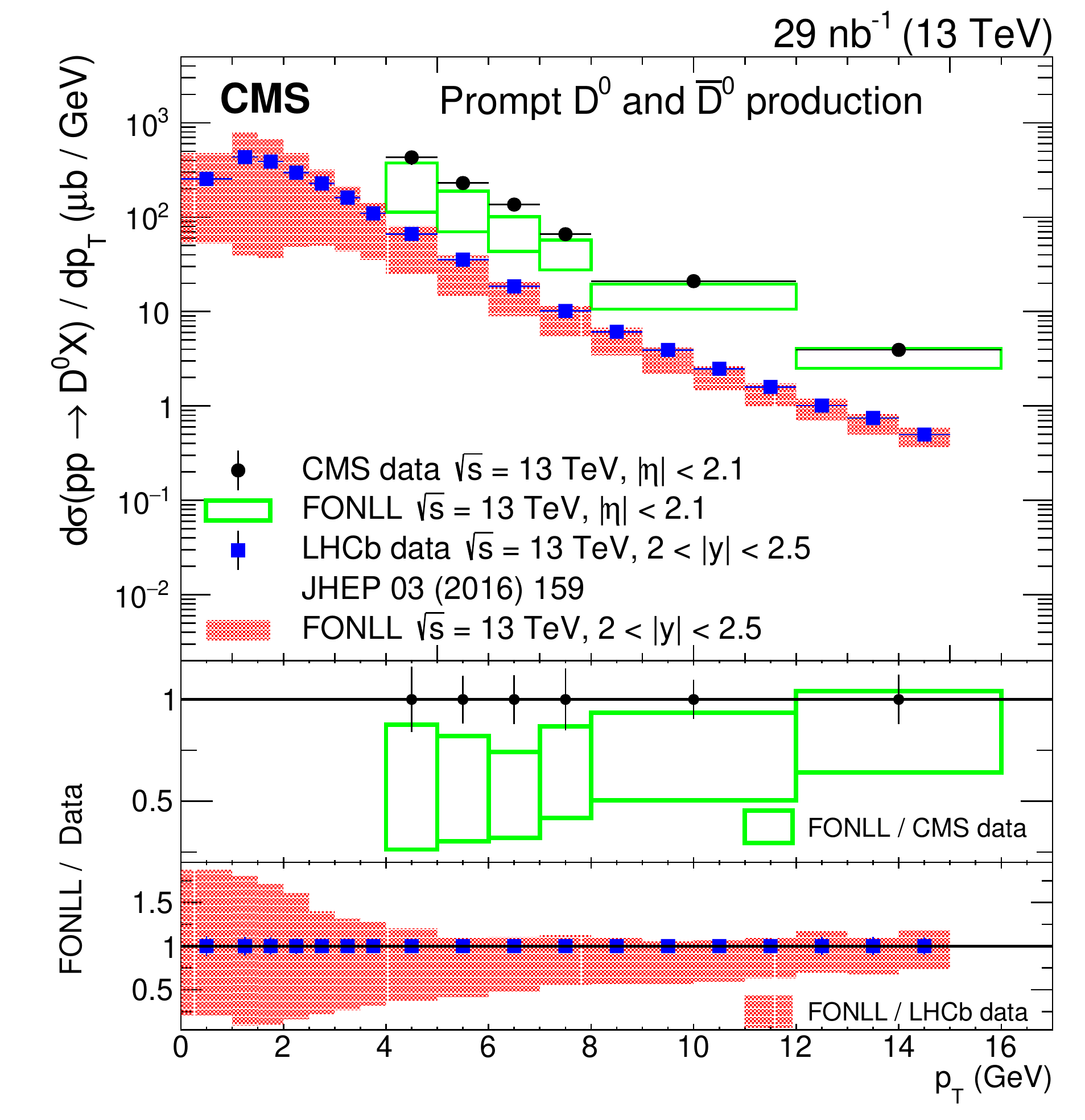}\\
\includegraphics[width=0.48\textwidth]{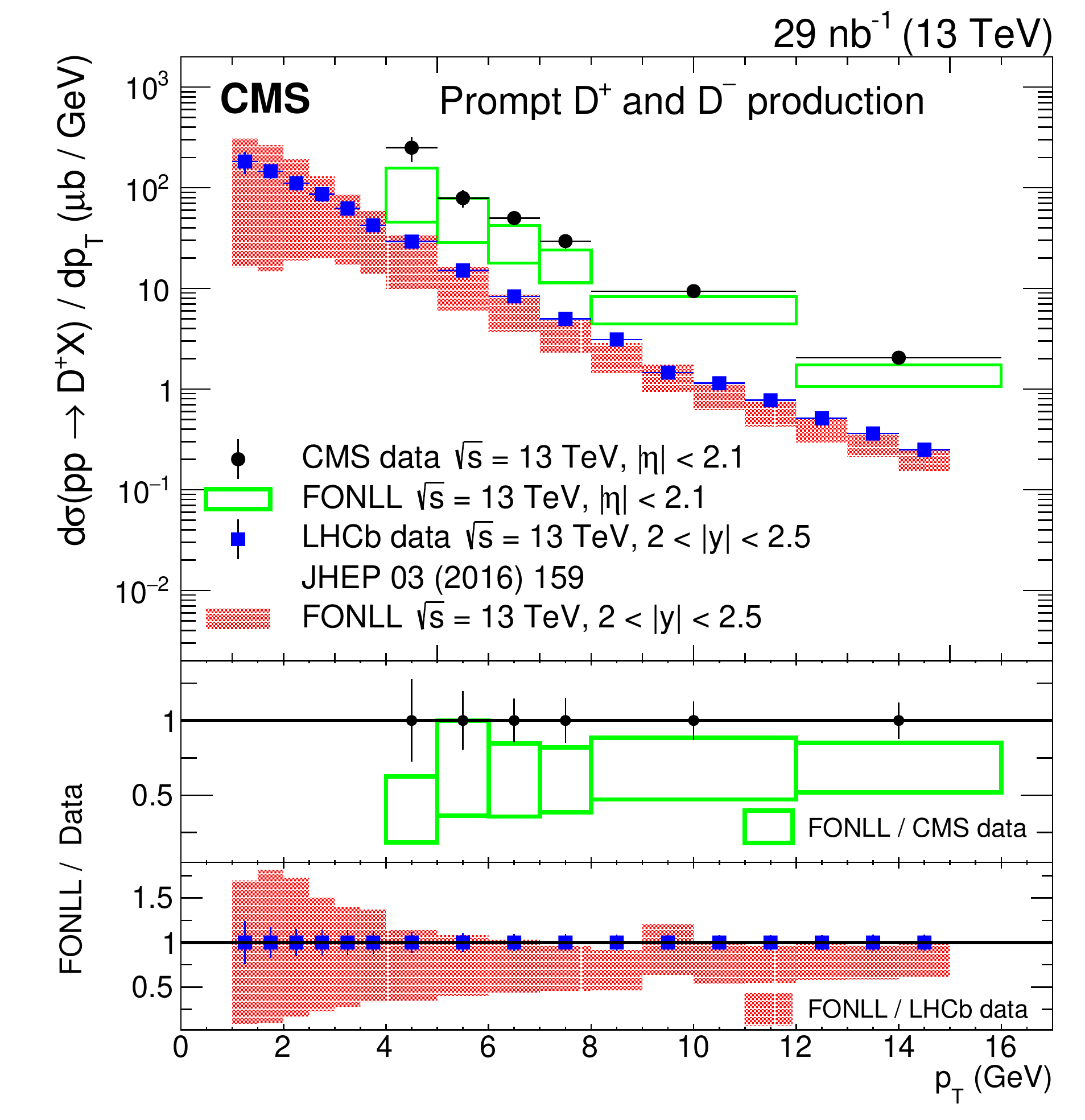}

\caption{Differential cross section $\rd\sigma/\rd\pt$ for prompt \PDstpm (upper left), \PDz$+$\PaDz (upper right) and \PDpm (lower) meson production at $\sqrt{s}= 13\TeV$ with $\pt < 16\GeV$ for CMS (black circles, this paper) for $\abs{\eta} < 2.1$  and LHCb \protect\cite{lhcb_13} (blue squares) for $2 < \abs{y} < 2.5$.
The corresponding FONLL predictions are shown by the unfilled and filled boxes, respectively.
To simplify the results representation, the equivalence between $\rd\sigma/\rd\pt$ for $2<\abs{y}<2.5$ and $\rd^2\sigma/\rd\pt dy$ for $2<y<2.5$, as in the original publication, has been used.
The vertical lines on the points give the total uncertainty in the data, and the horizontal lines show the bin widths.
The two lower panels in each plot give the ratios of the FONLL predictions to the CMS and LHCb data, shown by circles and squares, respectively.}
\label{LHCB_CMS}
\end{figure}

The only other measurement performed at $\sqrt{s}= 13\TeV$ for which the results are publicly available numerically (the results of \cite{alice_13} are only available in the figures) at $\sqrt{s}= 13\TeV$ comes from the LHCb Collaboration \cite{lhcb_13}.
Since the $\eta$ acceptances of the CMS and LHCb experiments differ, the two measurements are complementary and the results presented in this paper extend the reconstruction to a rapidity region not covered by LHCb.
The two measurements are compared in Fig.~\ref{LHCB_CMS} for the LHCb rapidity bin closest to the CMS fiducial region, together with the FONLL predictions. 
The CMS measurements are shown for $\pt < 16\GeV$ to allow a better comparison with the LHCb results. 
Again, both sets of measurements are in reasonable agreement with the FONLL predictions.

\clearpage
\section{Summary}

The differential cross sections $\rd\sigma/\rd\pt$ and $\rd\sigma/\rd\abs{\eta}$ for prompt charm meson (\PDstpm, \PDz (\PaDz), and \PDpm) production are measured in the transverse momentum range $4 < \pt < 100\GeV$ and pseudorapidity $\abs{\eta} < 2.1$, using data collected by the CMS experiment in proton-proton collisions in 2016 at $\sqrt{s} = 13\TeV$, corresponding to an integrated luminosity of 29\nbinv.
The charm mesons were identified with signal invariant mass peaks of high statistical significance. 
The contamination arising from nonprompt \PD mesons originating from \Pb hadron decays was removed using Monte Carlo event simulations, validated by measurements.

The measured cross section values are compared to predictions from a theoretical calculation and several different Monte Carlo generators. 
The agreement with the various models can be considered fair, but no single Monte Carlo simulation or theoretical prediction describes the data well over the entire kinematic range.
The measurements tend to favor a higher cross section than predicted by the FONLL calculations \cite{fonll, fonll2} and lower than estimated by the \PYTHIA event generators \cite{pythia6, pythia8}.
The cross section predictions from the different \PYTHIA tunes differ in both normalization and shape, which confirms that  the description of the data provided by the models is sensitive to further model improvements.
Overall, the best description is obtained by the upper edge of the FONLL uncertainty band, which could be taken as a reference prediction for background estimations for other processes, over the full kinematic range covered by all the LHC measurements. 
By confirming this finding in kinematic regions not previously covered, this measurement makes a contribution to the understanding of charm meson production in hadronic collisions, which is still dominated by large uncertainties in the present theoretical models.
 
\begin{acknowledgments}
  We congratulate our colleagues in the CERN accelerator departments for the excellent performance of the LHC and thank the technical and administrative staffs at CERN and at other CMS institutes for their contributions to the success of the CMS effort. In addition, we gratefully acknowledge the computing centers and personnel of the Worldwide LHC Computing Grid and other centers for delivering so effectively the computing infrastructure essential to our analyses. Finally, we acknowledge the enduring support for the construction and operation of the LHC, the CMS detector, and the supporting computing infrastructure provided by the following funding agencies: BMBWF and FWF (Austria); FNRS and FWO (Belgium); CNPq, CAPES, FAPERJ, FAPERGS, and FAPESP (Brazil); MES (Bulgaria); CERN; CAS, MoST, and NSFC (China); MINCIENCIAS (Colombia); MSES and CSF (Croatia); RIF (Cyprus); SENESCYT (Ecuador); MoER, ERC PUT and ERDF (Estonia); Academy of Finland, MEC, and HIP (Finland); CEA and CNRS/IN2P3 (France); BMBF, DFG, and HGF (Germany); GSRT (Greece); NKFIA (Hungary); DAE and DST (India); IPM (Iran); SFI (Ireland); INFN (Italy); MSIP and NRF (Republic of Korea); MES (Latvia); LAS (Lithuania); MOE and UM (Malaysia); BUAP, CINVESTAV, CONACYT, LNS, SEP, and UASLP-FAI (Mexico); MOS (Montenegro); MBIE (New Zealand); PAEC (Pakistan); MSHE and NSC (Poland); FCT (Portugal); JINR (Dubna); MON, RosAtom, RAS, RFBR, and NRC KI (Russia); MESTD (Serbia); SEIDI, CPAN, PCTI, and FEDER (Spain); MOSTR (Sri Lanka); Swiss Funding Agencies (Switzerland); MST (Taipei); ThEPCenter, IPST, STAR, and NSTDA (Thailand); TUBITAK and TAEK (Turkey); NASU (Ukraine); STFC (United Kingdom); DOE and NSF (USA).
   
  \hyphenation{Rachada-pisek} Individuals have received support from the Marie-Curie program and the European Research Council and Horizon 2020 Grant, contract Nos.\ 675440, 724704, 752730, 765710 and 824093 (European Union); the Leventis Foundation; the Alfred P.\ Sloan Foundation; the Alexander von Humboldt Foundation; the Belgian Federal Science Policy Office; the Fonds pour la Formation \`a la Recherche dans l'Industrie et dans l'Agriculture (FRIA-Belgium); the Agentschap voor Innovatie door Wetenschap en Technologie (IWT-Belgium); the F.R.S.-FNRS and FWO (Belgium) under the ``Excellence of Science -- EOS" -- be.h project n.\ 30820817; the Beijing Municipal Science \& Technology Commission, No. Z191100007219010; the Ministry of Education, Youth and Sports (MEYS) of the Czech Republic; the Deutsche Forschungsgemeinschaft (DFG), under Germany's Excellence Strategy -- EXC 2121 ``Quantum Universe" -- 390833306, and under project number 400140256 - GRK2497; the Lend\"ulet (``Momentum") Program and the J\'anos Bolyai Research Scholarship of the Hungarian Academy of Sciences, the New National Excellence Program \'UNKP, the NKFIA research grants 123842, 123959, 124845, 124850, 125105, 128713, 128786, and 129058 (Hungary); the Council of Science and Industrial Research, India; the Latvian Council of Science; the Ministry of Science and Higher Education and the National Science Center, contracts Opus 2014/15/B/ST2/03998 and 2015/19/B/ST2/02861 (Poland); the National Priorities Research Program by Qatar National Research Fund; the Ministry of Science and Higher Education, project no. 0723-2020-0041 (Russia); the Programa Estatal de Fomento de la Investigaci{\'o}n Cient{\'i}fica y T{\'e}cnica de Excelencia Mar\'{\i}a de Maeztu, grant MDM-2015-0509 and the Programa Severo Ochoa del Principado de Asturias; the Thalis and Aristeia programs cofinanced by EU-ESF and the Greek NSRF; the Rachadapisek Sompot Fund for Postdoctoral Fellowship, Chulalongkorn University and the Chulalongkorn Academic into Its 2nd Century Project Advancement Project (Thailand); the Kavli Foundation; the Nvidia Corporation; the SuperMicro Corporation; the Welch Foundation, contract C-1845; and the Weston Havens Foundation (USA).
\end{acknowledgments}

\bibliography{auto_generated}  
\cleardoublepage \appendix\section{The CMS Collaboration \label{app:collab}}\begin{sloppypar}\hyphenpenalty=5000\widowpenalty=500\clubpenalty=5000\vskip\cmsinstskip
\textbf{Yerevan Physics Institute, Yerevan, Armenia}\\*[0pt]
A.~Tumasyan
\vskip\cmsinstskip
\textbf{Institut f\"{u}r Hochenergiephysik, Wien, Austria}\\*[0pt]
W.~Adam, F.~Ambrogi, T.~Bergauer, M.~Dragicevic, J.~Er\"{o}, A.~Escalante~Del~Valle, R.~Fr\"{u}hwirth\cmsAuthorMark{1}, M.~Jeitler\cmsAuthorMark{1}, N.~Krammer, L.~Lechner, D.~Liko, T.~Madlener, I.~Mikulec, F.M.~Pitters, N.~Rad, J.~Schieck\cmsAuthorMark{1}, R.~Sch\"{o}fbeck, M.~Spanring, S.~Templ, W.~Waltenberger, C.-E.~Wulz\cmsAuthorMark{1}, M.~Zarucki
% ...
\vskip\cmsinstskip
% ...
1:  Also at TU Wien, Wien, Austria\\
% ...
\end{sloppypar}
\end{document}